\documentclass[prb,aps,amssymb,amsmath,twocolumn]{revtex4-2}
\usepackage[pdftex]{graphicx} 
\usepackage{orcidlink} 
\newcommand{\li}{\mathrm{Li}} 
\begin{document}

\title{Sound attenuation and velocity shift in antiferromagnetic spin-1/2 chains}
\author{Edmond Orignac\orcidlink{0000-0002-3405-9508}}
\affiliation{CNRS, Ens de Lyon, LPENSL, UMR5672,  F-69342 Lyon, France}
\author{Emeric Caprani\orcidlink{0009-0000-5069-4360}}
\affiliation{Universit\'e Lyon 1, CNRS, Institut Lumi\`ere Mati\`ere, UMR5306, F-69100, Villeurbanne, France}
\author{Roberta Citro\orcidlink{0000-0002-3896-4759}}
\affiliation{Physics Department ``E. R. Caianiello" and CNR-SPIN, Universit\'{a} degli Studi di Salerno, INFN, Gruppo Collegato di Salerno, 84084-Fisciano (Sa), Italy}
\date{\today}

\begin{abstract}  
  We investigate ultrasound attenuation and sound velocity shift in antiferromagnetic spin-$1/2$  XXZ chains in magnetic field. We relate the sound velocity shift to derivatives of the free energy with respect to exchange interactions, permitting its calculation with integrability techniques  at any temperature. Using bosonization, we predict the sound velocity shift exhibits a quadratic temperature correction at low temperatures in the Tomonaga-Luttinger liquid phase. Close to the fully polarized phase, a universal behavior associated with $z=2$ quantum criticality is found. In the Tomonaga-Luttinger liquid phase, ultrasound attenuation obeys a scaling law as a function of wavelength and temperature. An enhancement of attenuation is obtained 
  near the fully polarized phase. 

\end{abstract} 

\maketitle

\section{Introduction}
\label{sec:intro}


The sound velocity shift in spin chains is a classic probe of spin–lattice (magnetoelastic) coupling\cite{tani_1968,kawasaki_1970,tachiki_1974,pawlak_2009} and ultrasound measurements have been widely used to investigate quasi-one-dimensional quantum antiferromagnets \cite{poirier_2002,wolf_2004,chiatti_character_2008,yamaguchi_2011,poirier_2012,sergeicheva_2020,povarov_2024} through the coupling between acoustic phonons and spin degrees of freedom. From the theoretical side, both the renormalization of the sound velocity and the ultrasound absorption coefficient have been related to dynamical and static spin–spin correlation functions \cite{tani_1968,kawasaki_1970,tachiki_1974}, establishing a direct connection between measurable elastic properties and the underlying magnetic fluctuations. Strong effects are expected in proximity to magnetic or structural phase transitions \cite{pawlak_2009}, where the enhancement of spin correlations and the associated critical fluctuations lead to pronounced anomalies in the elastic response.

In integrable one-dimensional systems, such as the spin-1/2 XXZ chain, a more refined understanding has been achieved. In particular, Bethe Ansatz techniques \cite{takahashi_book_bethe} have been successfully employed to compute the ground-state contribution to the sound velocity shift\cite{tsyplyatyev_2017} and to relate it to thermodynamic derivatives of the free energy. More recently, extensions to finite-temperature regimes and dynamical response functions have been explored using integrability-based methods and effective field theory descriptions, further clarifying the role of low-energy excitations and Luttinger liquid physics in the renormalization of acoustic properties \cite{tsyplyatyev_2017, citro04_spinpeierls}.

Here, we derive the sound velocity shift and the ultrasound absorption coefficient in spin-1/2 XXZ chains by combining bosonization \cite{giamarchi_book_1d} with the quantum transfer matrix approach within a path-integral approach\cite{destri92_nostrings,destri95_tba_inteq,kluemper_heisenberg_thermo,kluemper_thermo_xxx,kluemper_xxz}. This approach allows us to access in a unified framework both low-energy universal contributions and full finite-temperature thermodynamics, overcoming limitations of purely field-theoretical or purely Bethe Ansatz ground-state treatments. In particular, we are able to capture the temperature evolution of magnetoelastic response functions across different regimes, from the Luttinger liquid phase to regimes dominated by thermal fluctuations, and to provide explicit expressions for both the elastic renormalization and attenuation in terms of finite-temperature correlation functions. Within this framework, we derive explicit expressions that connect magnetoelastic response functions to thermodynamic quantities. In particular, we show that in the Tomonaga–Luttinger liquid regime the velocity shift exhibits a universal low-temperature quadratic behavior in temperature, while the attenuation follows distinct power-law scalings depending on the relative magnitude of frequency and temperature. Near commensurate–incommensurate transitions, both quantities display singular behavior characteristic of enhanced critical fluctuations, whereas in the fully polarized regime the response is exponentially suppressed.

Beyond the low-energy regime, our approach also allows us to access the crossover to intermediate temperatures of order of the exchange coupling, where a full non-linear integral equation treatment becomes necessary. Overall, our results provide a unified description of elastic renormalization and sound attenuation in the XXZ chain, connecting field-theoretical and exact thermodynamic approaches, and clarifying the role of fluctuations and finite-temperature effects in magnetoelastic response functions.

The paper is organized as follows: In Sec. \ref{sec:shift-1chain} we first introduce the spin-1/2 XXZ model and its coupling to longitudinal lattice deformations relevant for magnetoelastic effects. In this Section we first derive the sound velocity shift following the Tachiki-Maekawa approach\cite{tachiki_1974}, treating first the case of the Heisenberg and XY chain, and then generalizing to the XXZ spin-1/2 chain.  The phonon renormalization is expressed in terms of thermodynamic derivatives of the spin-chain free energy. This allows us to establish general relations between the elastic response and spin correlation functions, which form the basis of the subsequent analysis.
In \ref{sec:attenuation-1chain} we analyze the sound attenuation.  The low-energy regime of the XXZ chain is discussed using bosonization and mapping the model onto a Tomonaga–Luttinger liquid\cite{Bouchoule2025}. Within this framework, we obtain explicit expressions for the temperature and momentum dependence of the sound velocity renormalization, clarifying its universal behavior at low temperatures. Finally we extend the analysis to the full finite-temperature regime using the quantum transfer matrix approach, and we derive the corresponding results for both the sound velocity shift and the ultrasound attenuation.

\section{Model and Hamiltonian}
We consider a single XXZ spin-1/2 chain in the presence of longitudinal acoustic phonons propagating along the chain. {In ultrasound experiments, the low energy and large number of acoustic phonons allow treating the phonons classically. However, a consistent derivation of the acoustic phonon frequency shift and attenuation using linear response theory requires that phonons are also treated quantum mechanically, the classical limit being taken at the end of the calculation.}   The Hamiltonian of the spin chain is 
\begin{equation}\label{eq:xxz-chain} 
  H_0=\sum_{j} J (S_{j}^x S_{j+1}^x + S_{j}^x S_{j+1}^x ) + J_z S_j^z S_{j+1}^z -h S_j^z, 
\end{equation}
with $h=g \mu_B B$, $B$ the magnetic induction, $\mu_B$ the Bohr magneton, $g$ the gyromagnetic ratio, $J$ and $J_z$ the exchange couplings. In the $XY$ spin chain, $J_z=0$ and in the Heisenberg spin chain, $J=J_z$. 
 The Hamiltonian for longitudinal phonons is 
\begin{equation}\label{eq:phonons}
 H_{ap}=\sum_k \omega_k \left(b^\dagger_k b_k + \frac 1 2\right),      
\end{equation}
where $b_k$ annihilates a phonon of momentum $k$, and $\omega_k = v_l |k|$ is the pulsation of a phonon of momentum $k$ with velocity $v_l$.  
The full Hamiltonian is $H=H_0+H_{ap}+H_1+H_2$,
where the spin-phonon interaction\cite{tachiki_1974,landau_statmech} is $H_1+H_2$. We have 
\begin{eqnarray}
&&H_1=\sum_k \frac{1}{\sqrt{2m N \omega_k}} (b_k + b_{-k}^\dagger)   U_k^{(1)} \\
&&H_2=\sum_{k,k'} \frac{1}{2m N \sqrt{\omega_k \omega_{k'}}} (b_k + b_{-k}^\dagger) (b^\dagger_{k'} + b_{-k'})  U_{k k'}^{(2)} 
\end{eqnarray}, 
where $m$ is the mass of atom, $N$ the number of sites in the chain, 
and the
spin phonon interaction matrix elements $U^{(1,2)}$ read\cite{tachiki_1974}
\begin{eqnarray}
  \label{eq:u1-decoupled} 
&&  U_{k}^{(1)} = -\sum_{j,\alpha=x,y,z} e^{i k ja}  (1-e^{i k a}) \frac{\partial J^\alpha}{\partial a} S_{j}^\alpha S_{j+1}^\alpha, \\
    \label{eq:u2-decoupled} 
&&  U_{kk'}^{(2)} = \frac 1 2 \sum_{j} e^{i
  (k-k')  ja  }  (1-e^{i k a}) (1-e^{-i k' 
  a}) \frac{\partial^2
  J^\alpha}{\partial a^2} {S}^\alpha_{j} \cdot
  {S}^\alpha_{j+1},  
\end{eqnarray}
where $k,k'$ are the wavevectors of the longitudinal phonons, $J^x=J^y=J$ and  $a$ is the lattice spacing.  
Note that our sign convention for $J$ is opposite to 
the one of Ref.~\onlinecite{tachiki_1974}, so we don't have minus signs. 
Eqs.~(\ref{eq:u1-decoupled})--(\ref{eq:u2-decoupled}) simply describe the change of spin-spin interaction resulting from atomic displacement caused by the longitudinal phonons.

\section{Sound velocity shift}
\label{sec:shift-1chain}

According to Ref.~\onlinecite{kawasaki_1970,tachiki_1974}, the shifts in sound pulsation are given by
\begin{eqnarray}
  \label{eq:shift-tachiki}
  \Delta \omega_{k} &=&  (\Delta \omega_{k})_1+ (\Delta \omega_{k})_2\nonumber \\
  \label{eq:shift1-tachiki}
  (\Delta \omega_{k})_1&=& -\frac{\beta}{2 m N \omega_{k}} \langle U_{k}^{(1)} U_{-k}^{(1)} \rangle \\
  \label{eq:shift2-tachiki}
   (\Delta \omega_{k})_2 &=& \frac {\langle U_{kk}^{(2)}\rangle}{2m N \omega_{k}}.    
\end{eqnarray}
However, Eq.~(\ref{eq:shift1-tachiki}) applies only to classical {spin} systems or to quantum systems in which $U_{k}^{(1)}$ is a conserved quantity. In a generic quantum system, we must replace Eq.~(\ref{eq:shift1-tachiki}) with the Matsubara linear response expression\cite{negele_orland} 
\begin{eqnarray}
 \label{eq:shift1-quantum}
  (\Delta \omega_{k})_1= -\frac{1}{2 m N \omega_{k}} \int_0^\beta d\tau \langle T_\tau U_{k}^{(1)}(\tau) U_{-k}^{(1)} \rangle.   
\end{eqnarray}
In the limit of high temperature, Eq.~(\ref{eq:shift1-quantum}) reduces to Eq.~(\ref{eq:shift1-tachiki}). 
We note that $(\Delta\omega_{k})_1<0$ but the sign of 
$(\Delta\omega_{k})_2$ is not known a priori.

If we consider $(\Delta\omega_{k})_2$, using the
definition~(\ref{eq:shift2-tachiki}) and Eq.~(\ref{eq:u2-decoupled}),
we arrive at the expression
\begin{eqnarray}\label{eq:shift2-chain-full}
  (\Delta\omega_{k})_2 = 4\sin^2(k a/2)   \sum_{\alpha=x,y,z} \frac{\partial^2 J^\alpha}{\partial a^2} \frac{\langle S^\alpha_0 \cdot S^\alpha_{1}\rangle}{2m\omega_k},  
\end{eqnarray}
showing that the frequency shift $(\Delta\omega_{k})_2$ is determined by equal-time spin-spin correlation functions. This generalizes the zero temperature contribution\cite{tsyplyatyev_2017} $\delta v_1$ to finite temperature.    Turning to the shift
$(\Delta\omega_{k})_1$, we obtain
\begin{eqnarray}\label{eq:shift1-chain-full}
   &&(\Delta\omega_{k})_1 =-\frac{1}{2m  \omega_{k}^0}  4\sin^2(k a/2)  \sum_{j,\alpha,\gamma} e^{i k j a} \int_0^\beta \frac{\partial J^\alpha}{\partial a} \frac{\partial J^\gamma}{\partial a} \times \nonumber \\ 
   && \langle T_\tau (S^\alpha_j S^\alpha_{j+1} - \langle S^\alpha_j S^\alpha_{j+1} \rangle) (\tau)  (S^\gamma_0  S^\gamma_{1} - \langle S^\gamma_0 S^\gamma_{1} \rangle) \rangle d\tau.    
\end{eqnarray}
The above result should be compared with $\delta v_2$ in Ref.~\cite{tsyplyatyev_2017}. Taking the zero temperature limit, and introducing a resolution of the identity, the second order perturbation theory for $\delta v_2$ is recovered. 
Acoustic phonons having a  wavelength  much larger than lattice spacing ($k a \ll 1$),  we will let $k\to 0$ in the sum of Eq.~(\ref{eq:shift1-chain-full}) and use energy and magnetization conservation  to simplify the resulting integral. 
\subsection{Case of Heisenberg and XY spin chains}
In the case of the XY or the Heisenberg spin chain, for any spin $S\ge 1/2$,  Eq.~(\ref{eq:shift1-chain-full}) can be expressed in terms of the Hamiltonian,  Eq.~(\ref{eq:xxz-chain}) yielding 
\begin{eqnarray}\label{eq:acoustic-1chain} 
  (\Delta\omega_k)_2 \simeq (k a)^2  \frac{\partial^2 J}{\partial a^2} \frac{\langle H_0 \rangle}{2m N J \omega_k }, \nonumber  \\
  (\Delta\omega_k)_1 \simeq - (k a)^2  \left(\frac{\partial J}{\partial a}\right)^2 \frac{\beta \langle (H_0-\langle H_0\rangle)^2\rangle }{2m N J^2 \omega_k}.   
\end{eqnarray}
since $\omega_k=v_l |k|$, these expressions give a relative sound velocity shift $\Delta \omega_k/\omega_k$ independent of frequency. 
If $F$ is the free energy of the spin chain, we have 
\begin{eqnarray}\label{eq:thermo-deriv}
\frac{\partial F}{\partial J} &=& \frac{\langle H_0 \rangle}{J}, \\ 
\frac{\partial^2 F}{\partial J^2} &=& -\beta \frac{\langle (H_0-\langle H_0\rangle)^2\rangle}{J^2},  
\end{eqnarray}
so the relative velocity shift reads  
\begin{eqnarray}\label{eq:rel-vshift}
 && \frac{(\Delta \omega_k)_1+(\Delta \omega_k)_2}{\omega_k} = \frac{(ka)^2}{2 m N \omega_k^2} \left[  \frac{\partial F}{\partial J} \frac{\partial^2 J}{\partial a^2} + \frac{\partial^2 F}{\partial J^2} \left(\frac{\partial J}{\partial a}\right)^2\right]\nonumber \\ 
 &&= \frac{a^2}{2 m N v_l^2 } \frac{\partial^2 F}{\partial a^2}, 
\end{eqnarray}
which is simply proportional to the relative contribution of the spin chain free energy to the compressibility\cite{landau_elasticity}. 
In the absence of external magnetic field, $(\Delta\omega_k)_1$ is expressed using the specific heat
$C_V =\partial \langle H \rangle /\partial T$, yielding
\begin{eqnarray}
  \label{eq:shift-1chain}
  \frac{\Delta \omega_{k}}{\omega_k} =\frac{a^2}{2m N v_l^2} \left[\frac{1}{J} \frac{\partial^2 J}{\partial a^2} \langle H \rangle - \left(\frac{1}{J} \frac{\partial J}{\partial a} \right)^2 T C_V \right]. 
\end{eqnarray}
In the XY and in the Heisenberg chain, once the dependence of exchange interaction on interatomic distance is
known, the velocity shift is determined from internal energy and specific heat.
 If we restrict ourselves to low
temperatures, $k_B T \ll J$, and spin $1/2$,  we can use bosonization\cite{giamarchi_book_1d}
to estimate the free energy of the spin chain.  We have
\begin{eqnarray}\label{eq:free-energy-bosonized}
  \frac{F} {Na} =\frac{E_{GS}}{Na} - \frac{\pi (k_B T)^2}{6 u},  
\end{eqnarray}
where $u$ is the velocity of spin excitations. In the 
spin-1/2 Heisenberg chain, it is 
$u = \frac \pi 2 J a$, whereas $u=Ja$ in the XY spin chain.  
Using these results, we find
that in the Heisenberg chain, for longitudinal acoustic phonons with dispersion
$\omega_k = v_l k$ , the velocity is shifted by
\begin{eqnarray}
  \frac{\Delta v_l(T)}{v_l} = \frac{\Delta v_l (T=0)}{v_l} - \frac{a^2 T^2 }{6 m v_l^2 } \frac{\partial^2} {\partial a^2} \left(\frac{1}{J(a)} \right) ,  
\end{eqnarray}
where $(\Delta v_l)_{T=0}$ has already been calculated
 using integrability techniques\cite{tsyplyatyev_2017}. A similar result obtains in the XY chain. 
 When the magnetic field is turned on, the thermodynamic derivatives in Eq.~(\ref{eq:thermo-deriv}) become a bit more involved.  
In both the Heisenberg or XY spin chain, for any spin $S\ge 1/2$, the free energy takes the scaling form   
\begin{equation}
    F=k_B T f\left(\frac{k_B T}{J}, \frac h {J} \right), 
\end{equation}
and using homogeneity, we find
\begin{eqnarray}\label{eq:thermo-field-deriv-1}
\frac{\partial F}{\partial J} &=& \frac{E +h M}{J} \\ 
\label{eq:thermo-field-deriv-2}
\frac{\partial^2 F}{\partial^2 J} &=& -\frac{1}{J^2} \left[T C_V + 2h T \frac{\partial M}{\partial T} + h^2 \frac{\partial M}{\partial h} \right],  
\end{eqnarray}
so the sound velocity shift now depends not just on internal energy $E$, and specific heat $C_v$ but also on magnetization $M$, magnetic susceptibility and magnetocaloric effect.\cite{zhitomirsky04_magnetocaloric} 
In the ground state, when the magnetization becomes fully polarized, it exhibits a cusp singularity while the magnetic susceptibility shows a divergence\cite{japaridze_cic_transition,pokrovsky_talapov_prl,schulz_cic2d,chitra_spinchains_field,cabra_instabilityLL,orignac_2005_magnetostriction}. At finite temperature this singular behavior turns into a quantum critical point\cite{sachdev_qaf_magfield,blosser_2017,blosser_2018} with dynamical exponent $z=2$, where the magnetization follows a universal scaling law\cite{maeda_universal_2007}. 
 Within bosonization, for a spin-1/2 chain in the vicinity of a fully polarized state\cite{chitra_spinchains_field}, $K(h) \to 1$ and $u(h) \sim |h-h_c(J)|^{1/2}$ yielding a divergent correction to the phonon velocity at low temperature. Experiments on a Cu(II) coordination polymer have indeed revealed an anomaly in the longitudinal sound velocity near the saturation field\cite{wolf_2004} and  phenomenological formula was used to fit the measurements. Near saturation, it predicted a shift of sound velocity proportional to $\partial M/\partial h$, in agreement with Eq.~(\ref{eq:thermo-field-deriv-2}) and the divergent behavior of the susceptibility. 
Those very low temperature predictions, obtained in the framework of bosonization,  can be extended to general temperature
temperature. 
\subsubsection{XY chain case}
In the case of the XY spin chain, using the Jordan-Wigner transformation\cite{jordan_transformation} the exact free energy has been obtained analytically\cite{katsura_1962}. We have\cite{caprani_absorption_2023}
\begin{eqnarray}
\label{eq:xy-1stderiv}
\frac 1 {Na} \frac{\partial F} {\partial J} = \int_{-\frac{\pi} a}^{\frac{\pi} a} \frac{dk}{2\pi} \frac{\cos (ka)}{e^{(J\cos (ka) -h)/T}+1}, \\  
\label{eq:xy-2ndderiv}
\frac 1 {Na} \frac{\partial^2 F} {\partial J^2} = -\int_{-\frac{\pi} a}^{\frac{\pi} a} \frac{dk}{2\pi} \frac{\cos^2 (ka)}{4T\cosh^2\left(\frac{J\cos (ka) -h}{2T}\right)}. 
\end{eqnarray}
Under a particle-hole transformation, one can turn $h \to -h$, $\langle S_j^z \rangle  \to -\langle S_j^z\rangle$ so it is only necessary to consider the case of $h\to -J$, $\langle S_j^z \rangle \to -1/2$. In the limit $T\to 0$, we have 
\begin{eqnarray}
\frac{1}{Na} \frac{\partial F} {\partial J} \to -\frac 1 {\pi a} \sqrt{1 -\frac{h^2}{J^2}} \theta (J-|h|) \nonumber \\ 
\frac{1}{Na} \frac{\partial F} {\partial J} \to -\frac {h^2} {\pi J^2  a \sqrt{1 -\frac{h^2}{J^2}} } \theta (J-|h|), 
\end{eqnarray}
 {This would result in a divergent sound velocity shift} for $h=\pm J$, that is when the magnetization saturates. {In reality, such divergence simply indicates that the rigid lattice approximation breaks down when the spin chain magnetization is close to saturation.\cite{mattis63_magnetostriction,orignac_2005_magnetostriction,derzhko_2013} When the chain gets magnetized, it can contract to enhance its magnetic energy, until this is balanced by the elastic energy, while in the fully polarized case, no magnetic energy can be gained. This results in a discontinuity of the magnetization and of the lattice parameter at saturation, \textit{i. e.} a first order transition. When the temperature is high enough, the crossover between the fully polarized state and the partially magnetized state is sufficiently gradual to prevent a discontinuity. In that limit, perturbation theory is fully justified.}  To consider the behavior at finite temperature, 
close to saturation, only the low-energy spectrum is needed. 
When $h\to -J$, we can approximate $J\cos(ka)-h \simeq J (|k|a -\pi)^2/2 -(J+h)$ in Eqs.~(\ref{eq:xy-1stderiv})--(\ref{eq:xy-2ndderiv}), so that 
\begin{eqnarray}\label{eq:xy-univ}
\frac 1 {Na} \frac{\partial F} {\partial J} \simeq \sqrt{\frac{T}{2\pi J a^2}} \li_{1/2}\left(-e^{\frac{J+h} T}\right) + O(T^{3/2}) , \nonumber \\
\frac 1 {Na} \frac{\partial^2 F} {\partial J^2} \simeq \frac{1}{\sqrt{2\pi J Ta^2}} \li_{-1/2}\left(-e^{\frac{J+h} T}\right)+O(T^{1/2}),
\end{eqnarray}
where $\li$ is the polylogarithm function\cite{olver2010nist}. The derivatives are plotted on. 
\begin{figure}
    \centering
    \includegraphics[width=\linewidth]{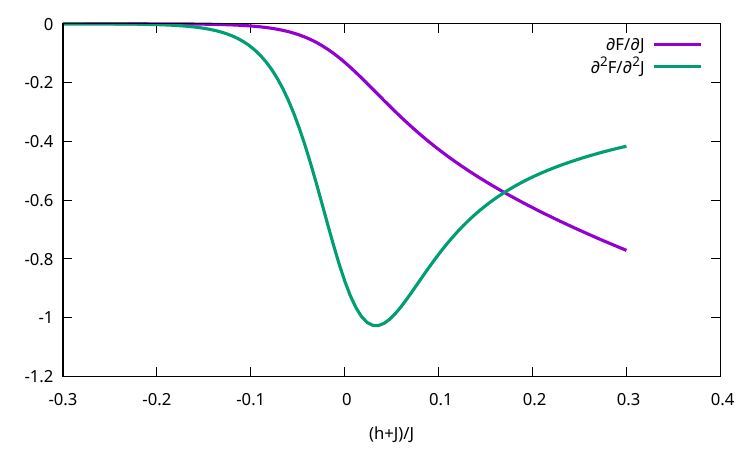}
    \caption{First and second derivatives of the free energy of the XY spin chain in the vicinity of $h=-J$ from Eq.~(\ref{eq:xy-univ}) for $T=0.03J$. The  first derivative is going to zero, and would present a cusp at $T=0$ while the second derivative presents a strong dip near the critical field, that becomes a divergence for $T\to 0$. As a result, the sound velocity shift shows a dip near saturation. }
    \label{fig:xy-approx}
\end{figure}

Those  expressions are a consequence\cite{maeda_universal_2007,guan_critical_2014,zheludev_2020} of the universal form of the free energy in the vicinity of of a commensurate-incommensurate transition\cite{sachdev_qaf_magfield,chitra_spinchains_field} of a gapped spin chain. Indeed, in the vicinity of the commensurate-incommensurate transition, the low energy physics of the gapped spin chain is described by free fermions\cite{japaridze_cic_transition,pokrovsky_talapov_prl,schulz_cic2d} with a quadratic dispersion and a varying chemical potential, just like the XY model near saturation. We thus expect Eqs.~(\ref{eq:xy-univ}) to be also applicable to velocity shifts for magnetic field close to the end of a magnetization plateau.\cite{oshikawa_plateaus,cabra_magnetization_plateaus} Indeed, experiments in DTN\cite{chiatti_character_2008,povarov_2024} have observed dips of the sound velocity in the vicinity of the Haldane-gap to Tomonaga-Luttinger phase transition.
At the critical point, $h=-J$, Eq.~(\ref{eq:xy-univ}) predicts $(\Delta \omega_k) \sim \mathrm{Li}_{-1/2}(-1)(2\pi J T)^{-1/2} + O(T^{1/2})$ with $\mathrm{Li}_{-1/2}(-1)=-0.3801048$. 
Below the critical point, $h<-J$, it predicts $(\Delta \omega_k) \sim e^{-|J+h|/T} $ and above the critical point $(\Delta \omega)_k \propto (2\pi^2 J(J+h))^{-1/2} + \pi T^2/(8  \sqrt{2J (h+J)^5}) + O(T^4)$. The latter result is in agreement with bosonization. Note that at low {enough} temperature\cite{mattis63_magnetostriction,orignac_2005_magnetostriction,derzhko_2013}, magnetostriction can turn the quantum critical point into a first order point, where the elastic constants become hysteretic. The dip in the velocity shift {is} then replaced by a discontinuity. {To properly describe that effect, it is necessary to go beyond the rigid lattice approximation and calculate self-consistently the lattice spacing in equilibrium taking into account the variation of the exchange interaction with the lattice spacing. Then, sound velocity shift and attenuation are obtained from Eqs.~(\ref{eq:shift1-chain-full})--(\ref{eq:att-matsubara}) in which the selfconsistently determined exchange interaction is used.}
\subsubsection{Case of the Heisenberg chain}
\begin{figure}
    \centering
    \includegraphics[width=\linewidth]{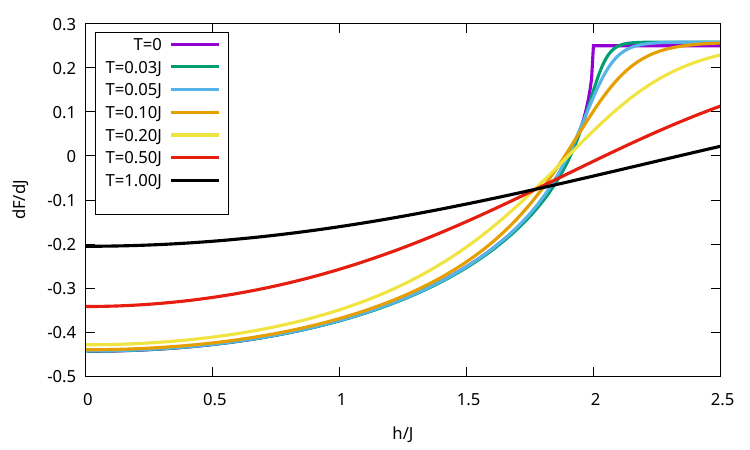}
    \caption{First derivative of the free energy of the Heisenberg spin-1/2 chain with respect to the exchange interaction as a function of the magnetic field for varying temperatures. In the ground state, the derivative goes to a constant after the critical field $h=2J$. Just below the critial field, it shows a cusp of the form $1/4-\mathcal{C}(2J-h)^{1/2}$. At finite temperature, the cusp is rounded off, and becomes unobservable for $T>J/2$. } 
    \label{fig:nlie-1st-deriv}
\end{figure}

\begin{figure}[b]
    \centering
    \includegraphics[width=\linewidth]{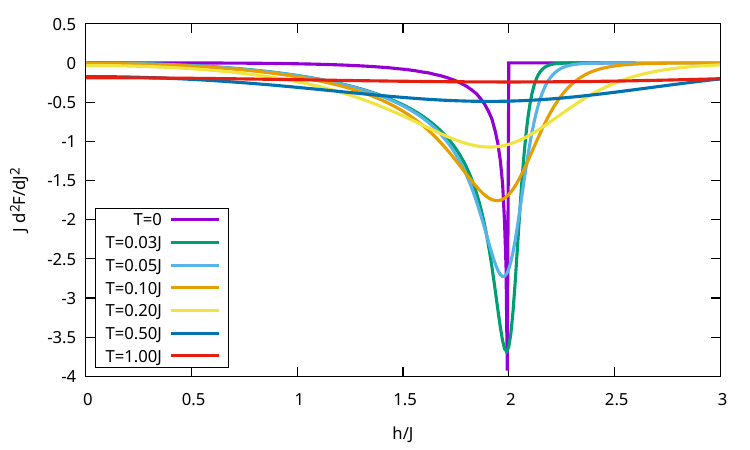}
    \caption{Second derivative of the free energy of the Heisenberg spin-1/2 chain with respect to the exchange interaction as a function of the magnetic field for varying temperatures. In the ground state, the second derivative exhibits a divergence $\sim (2J-h)^{-1/2}$. As temperature is increased, the divergence is replaced by a dip that becomes broader and shallower as temperature is increased. At temperatures larger than $J/2$, only a minimum can be observed.}
    \label{fig:nlie-2nd-deriv}
\end{figure}

In the case of the Heisenberg spin chain, the free energy at any temperature can be calculated with the thermodynamic Bethe
Ansatz\cite{takahashi_book_bethe} or the nonlinear integral equation
following the development of the path integral formulation of
the quantum transfer matrix approach\cite{destri97_tba_inteq,kluemper_thermo_xxx}. 
With the latter\cite{kluemper_thermo_xxx}, the first and second derivatives of the free energy in the Heisenberg chain with respect to exchange interaction can be obtained numerically. The details are explained in the Appendix\ref{section:NLE}. Their behavior as a function of temperature and magnetic field is represented on Figs.~\ref{fig:nlie-1st-deriv} and~\ref{fig:nlie-2nd-deriv}. Near the saturation field, a cusp-like singularity is seen in the first derivative, while the second derivative has dip, as expected from bosonization\cite{tsyplyatyev_2017}. As a result, the sound velocity near saturation will show a dip as found in experiment\cite{wolf_2004}. In fact, it has been shown\cite{he_2017,breunig_quantum_2017} that in the vicinity of the saturation, the asymptotic expansion of the free energy of the Heisenberg spin-1/2 chain takes the universal form\cite{sachdev_qaf_magfield,maeda_universal_2007,zheludev_2020} expected in the vicinity of a commensurate-incommensurate transition. So the analytic expressions~(\ref{eq:xy-univ}) derived for the XY chain are also applicable to the Heisenberg chain near saturation. Qualitatively, this can be seen by comparing the plots of the derivatives of the free energy close to saturation for the Heisenberg chain (Figs.~\ref{fig:nlie-1st-deriv} and~\ref{fig:nlie-2nd-deriv}) with those for the XY chain (Fig.~\ref{fig:xy-approx}).  Away from saturation, the full Bethe Ansatz solution is necessary for quantitative calculations. In that regime, the second derivative is negligible, and the first derivative dominates the frequency shift. The largest shift is expected at zero magnetic field. 


\subsection{Case of the general XXZ spin-1/2 chain}
When considering the general XXZ spin-1/2 chain, we must first distinguish $|J_z|>J$ and $|J_z|<J$. In the former case, a gap $\Delta$ above the ground state is present\cite{yang_xxz}, and the velocity shift is $O(e^{-\Delta/T}$ at low temperature. In the latter case, a Tomonaga-Luttinger liquid phase is present,\cite{nijs_equivalence} and a significant velocity shift is possible even for $T\ll J$ as we have seen in the case of the XY and Heisenberg spin-1/2 chain. However, in contrast to the case of Heisenberg and XXZ spin-1/2 chains, the derivative of the Hamiltonian $H_0$ with respect to lattice spacing is not anymore proportional to $H_0$ and as a consequence is not a conserved quantity. 
Therefore, in Eq.~(\ref{eq:shift1-chain-full}) the time dependence of the operators under the integral sign must be kept. We have to use the more general relations 
\begin{eqnarray}\label{eq:xxz-partderiv1}
&&\frac{\partial^2 F}{\partial^2 J} = -\int_0^\beta d\tau \left\langle T_\tau \left(\sum_{j,\alpha=x,y} S_j^\alpha S_{j+1}^\alpha - \langle S_j^\alpha S_{j+1}^\alpha \rangle  \right)(\tau) \right. \nonumber \\ 
&&\times\left.\left(\sum_{\gamma=x,y} S_0^\gamma S_{1}^\gamma - \langle S_0^\gamma S_{1}^\gamma \rangle  \right)(0)\right\rangle, \\ 
\label{eq:xxz-partderiv2}
&&\frac{\partial^2 F}{\partial^2 J^z} = -\int_0^\beta d\tau \left\langle T_\tau \left(\sum_{j} S_j^z S_{j+1}^z - \langle S_j^z S_{j+1}^z \rangle  \right) (\tau)\right. \nonumber \\ 
&&\times\left. \left( S_0^z S_{1}^z - \langle S_0^z S_{1}^z \rangle  \right)(0)\right\rangle, \\ 
\label{eq:xxz-partderiv3}
&&\frac{\partial^2 F}{\partial J \partial J_z} = -\int_0^\beta d\tau \left\langle T_\tau \left(\sum_{j,\alpha=x,y} S_j^\alpha S_{j+1}^\alpha - \langle S_j^\alpha S_{j+1}^\alpha \rangle  \right)(\tau) \right. \nonumber \\ 
&&\times\left. \left( S_0^z S_{1}^z - \langle S_0^z S_{1}^z \rangle  \right)(0)\right\rangle, \\
&=& -\int_0^\beta d\tau\left\langle T_\tau \left(\sum_{j,z} S_j^z S_{j+1}^z - \langle S_j^z S_{j+1}^z \rangle  \right)(\tau) \right. \nonumber \\ 
&&\times\left. \left(\sum_{\gamma=x,y} S_0^\gamma S_{1}^\gamma - \langle S_0^\gamma S_{1}^\gamma \rangle  \right)(0)\right\rangle, \\ 
\end{eqnarray}
showing that Eq.~(\ref{eq:rel-vshift}) remains valid, and allowing us to obtain the velocity shift using the exact solution of the XXZ spin-1/2 chain\cite{takahashi_book_bethe,kluemper_xxz}. However, since the free energy now takes the form $F=k_B T f(k_B T/J,h/J,J_z/J)$, it is not possible to express the partial derivatives (~\ref{eq:xxz-partderiv1})--(~\ref{eq:xxz-partderiv3}) using only the specific heat, the magnetocaloric effect, and the magnetic susceptibility as we could do in the case of the XY or the Heisenberg chain since the ratio $J_z/J$ can vary with the lattice spacing.  At low temperature, and in the absence of magnetic field, the velocity shift of the acoustic phonons is obtained from the spinon velocity\cite{giamarchi_book_1d} as
\begin{equation}
    \frac{\Delta v_l}{v_l} (T) = \frac{\Delta v_l}{v_l} (T=0) -\frac{a^2 T^2}{6m v_l^2} \frac{\partial^2}{\partial a^2} \left(\frac{\arccos (J_z/J)}{\sqrt{J^2-J_z^2}}\right),  
\end{equation}
where both $J$ and $J_z$ depend on the lattice spacing $a$. In the presence of a magnetic field, the velocity shift is obtained from the derivative of Eq.~(\ref{eq:free-energy-bosonized}), where the spinon velocity $u$ is determined numerically\cite{haldane_xxzchain}. 
For larger temperatures, the thermodynamic Bethe Ansatz\cite{takahashi_book_bethe} or the non-linear integral equation\cite{kluemper_xxz} can be used to calculate numerically the velocity shift, using slightly more general equations\cite{destri95_tba_inteq} than those of App.~\ref{section:NLE}. 
\section{Sound attenuation}
\label{sec:attenuation-1chain}
The ultrasound attenuation coefficient is\cite{tani_1968,tachiki_1974}
\begin{eqnarray}
  \label{eq:attenuation}
  \alpha_{k} = \mathrm{Re}\int_0^{+\infty} \frac{dt}{v_l} \frac{(f_{k}(t),f_{k}(0))}{(b_k,b_k)}  e^{-i \omega_{k} t},   
\end{eqnarray}
where the scalar product is defined as
\begin{eqnarray}
  (A,B)=\int_0^\beta d\lambda \frac{\mathrm{tr} (e^{-\beta H} e^{\lambda H} A e^{-\lambda H} B^\dagger)}{\mathrm{tr}(e^{-\beta H}) } - \beta \langle A\rangle  \langle A^\dagger\rangle,  
\end{eqnarray}
$v_l$ is the longitudinal sound velocity,  and the force operator is $f_k=-i [b_k,(H_1+H_2)]$. As shown in the appendix,
the expression~(\ref{eq:attenuation}) can be brought to the more
familiar form of the imaginary part of the response function
\begin{eqnarray}
  \label{eq:attenuation-alt}
  \alpha_{k} = \frac 1 {v_l} \mathrm{Im} \int_0^{+\infty} \langle [f_k(t),f^\dagger_k(0)]\rangle e^{-i \omega_{k}t} dt.   
\end{eqnarray}
which is convenient for perturbative calculations with the Matsubara formalism.\cite{negele_orland}

\begin{figure}
    \centering
    \includegraphics[width=\linewidth]{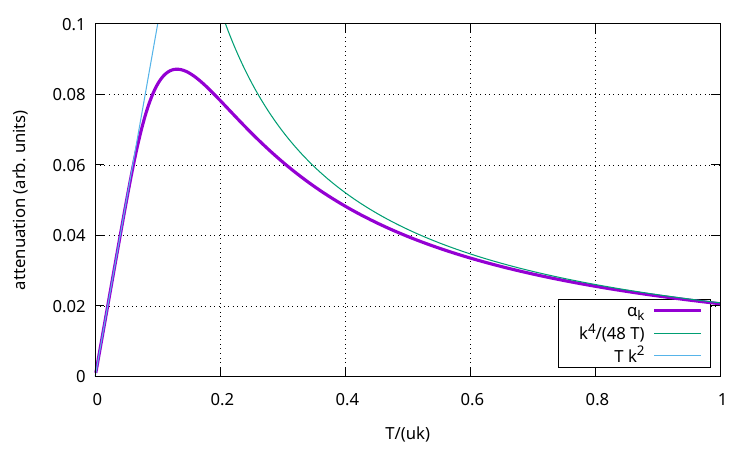}
    \caption{Plot of attenuation in the XXZ chain as a function of temperature for a fixed value of the wavevector $k$. At low temperatures, $T\ll uk$, the attenuation is linear in temperature. At high temperatures, it is inversely proportional to temperature. The attenuation reaches its maximum for $T=0.1308 uk$. }
    \label{fig:attenuation-T}
\end{figure}

In the case of sound attenuation, the random force operator
is\cite{tachiki_1974} at first order $[f_k,H_1]$  
\begin{eqnarray}
  f_k=- \left(\frac{1}{2 mN  \omega_{k}}\right)^{1/2} \sum_{j} e^{i k j a }  (1-e^{i k a})  \frac{\partial J^\alpha}{\partial a} S^\alpha_{j} \S^\alpha_{j+1},  
\end{eqnarray}
where we have dropped the $-i$ phase factor since it is cancelled by the
hermitian conjugate in the calculation of the response function. 
Using Eq.~(\ref{attenuation-response}) from App.~\ref{app:response},
and the Matsubara formalism\cite{mahan_book}, we end up having to
calculate
\begin{eqnarray}\label{eq:att-matsubara} 
 && \alpha_M(i\omega_n) = \frac{1}{2 m N \omega_{k} v_l} \int_0^{\beta}  d\tau e^{i\omega_n \tau} \langle T_\tau U_{-k}^{(1)}(\tau) U_{k}^{(1)}(0)\rangle \nonumber \\
&&  = \frac{4 \sin^2(ka/2)}{2 m \omega_k v_l}   \sum_{j,\alpha,\gamma}  \int_0^{\beta} d\tau \frac{\partial J^\alpha}{\partial a} \frac{\partial J^\gamma}{\partial a}  \nonumber \\ 
&& \times  \langle T_\tau (S^\alpha_j S^\alpha_{j+1}) (\tau) (S^\gamma_0 \cdot S^\gamma_{1}) (0)\rangle e^{-i(k j a -\omega_n \tau)},  
\end{eqnarray}
and take the analytic continuation $i\omega_n \to -\omega_{k} + i 0_+$ to obtain $\alpha_k$. Let us first consider the case without applied magnetic field. 
We can rewrite
\begin{eqnarray}\label{eq:attenuation-op1}
\sum_\alpha  \frac{\partial J^\alpha}{\partial a} S^\alpha_j S^\alpha_{j+1} &=& \frac {1}{J} \frac{\partial J}{\partial a} [J(S_j^x S_{j+1}^x + S_j^{y} S_{j+1}^y) + J^z S^z_{j} S_{j+1}^z ] \nonumber \\ 
&&   + \left(\frac{\partial J^z}{\partial a} -\frac{J^z}{J}\right) \frac{\partial J}{\partial a} S_j^z S_{j+1}^z,      
\end{eqnarray}
  In the $XY$ chain, since $J_z=0$, and in the Heisenberg chain, since $J_z=J$ the attenuation depends only on the correlator of the Hamiltonian density.  To be precise, we must distinguish genuine XY and Heisenberg spin chains, in which the condition $J^z=0$ or $J=0$ is satisfied independently of lattice spacing, from accidental XY or Heisenberg spin chains, that is XXZ spin chains in which a $XY$ or Heisenberg is realized for a particular value of the lattice spacing. Only in the former does the attenuation depend only on Hamiltonian density.  
Now, we turn to the evaluation of ultrasound attenuation in a generic XXZ chain using bosonization\cite{giamarchi_book_1d}. We can rewrite Eq.~(\ref{eq:attenuation-op1}) 
\begin{eqnarray}
 && \sum_\alpha  \frac{\partial J^\alpha}{\partial a} S^\alpha_j S^\alpha_{j+1} =\frac{a}{J} \frac{\partial J}{\partial a} \left[\frac{uK}{2\pi}(\pi \Pi)^2 + \frac{u}{2\pi K} (\partial_x \phi)^2 \right] \nonumber \\ 
&& + \frac{a^2}{\pi^2} \left(\frac{\partial J^z}{\partial a} -\frac{J^z}{J} \frac{\partial J}{\partial a}\right) (\partial_x \phi)^2 + \mathrm{staggered},     
\end{eqnarray}
where the first term comes from the density of Hamiltonian, and the second term from $S_j^z S_{j+1}^z$. We have not written the staggered terms because they have a wavevector $\frac{\pi}{a}$ much larger than the wavevector of any acoustic phonon. We have dropped from the Hamiltonian density a term $\cos (4\phi)$ that is present in the absence of magnetic field, but irrelevant for $J_z<J$. To calculate the correlator, Eq.~(\ref{eq:att-matsubara}), we introduce the chiral fields 
\begin{eqnarray}
\phi=\frac{\sqrt{K}}{2} (\varphi_L +\varphi_R), \\
\theta=\frac{\sqrt{K}^{-1}}{2} (\varphi_L -\varphi_R), 
\end{eqnarray}
and rewrite 
\begin{eqnarray}
&& \sum_\alpha  \frac{\partial J^\alpha}{\partial a} S^\alpha_j S^\alpha_{j+1} =\left[\frac{a u }{4\pi J} \frac{\partial J}{\partial a} \frac{a^2 K}{4 \pi^2} \left(\frac{\partial J^z}{\partial a} -\frac{J^z}{J} \frac{\partial J}{\partial a}\right) \right] \nonumber \\ 
&& \times \left[ (\partial_x \varphi_R)^2 + (\partial_x \varphi_L)^2 \right] + \frac{a^2 K}{\pi^2} \left(\frac{\partial J^z}{\partial a} -\frac{J^z}{J} \frac{\partial J}{\partial a}\right) \partial_x \varphi_R \partial_x \varphi_L,     
\end{eqnarray}
since the correlator $\langle \varphi_R \varphi_L \rangle =0$, the cross correlations vanish, and we only have to consider correlators $\langle (\partial_x\varphi_R)^2(x,\tau) (\partial_x\varphi_R)^2(0,0) \rangle  $, $\langle (\partial_x \varphi_L)^2(x,\tau) (\partial_x \varphi_L)^2(0,0) \rangle $ and  $\langle \partial_x \varphi_R^(x,\tau) \partial_x \varphi_R(0,0) \rangle  \langle \partial_x \varphi_L(x,\tau) \partial_x \varphi_L(0,0) \rangle$. The first correlator depends only on $x+i u \tau$, the second depends only on $x-i u \tau$ and the third depends on both. After Fourier transformation and analytic continuation, the contribution of the first correlator to the attenuation turns out to be proportional to $\delta(\omega_k - u k)$ while the one of the second correlator turns out to be proportional to $\delta(\omega_k + uk)$. Since in a spin chain $u \gg v_l$ the contribution of those terms is vanishing. Only the third correlator contributes to ultrasound attenuation. 
We have 
\begin{eqnarray}\label{eq:correlator-marginal}
&&\langle T_\tau \varphi_R(x,\tau) \varphi_R(0,0) \rangle \langle T_\tau \varphi_R(x,\tau) \varphi_R(0,0) \rangle \nonumber\\ 
&&= \left(\frac{2 \pi^2 }{(\beta u)^2 \left[\cosh \left(\frac{2\pi x}{\beta u}\right) - \cos \left(\frac{2\pi a}{\beta u}\right) \cos \left(\frac{2\pi \tau }{\beta} \right) \right]} \right)^2,       
\end{eqnarray}
where the factor $\cos [2\pi\alpha/(\beta u)]$ has been introduced to provide a short distance cutoff. 
After Fourier transformation and analytic continuation\cite{giamarchi_book_1d}, 
\begin{eqnarray}
&& \alpha_k = \frac{k^3 a^5 K^2 (u^2 -v_l^2) }{8m \pi^3 m v_l^2 u^3}  \left(\frac{\partial J^z}{\partial a} -\frac{J^z}{J} \frac{\partial J}{\partial a}\right)^2 \\ 
&&\times \left[\frac{\pi}{2 \tanh \frac{\beta (u+v_l) k}{4}} - \frac{\pi}{2 \tanh \frac{\beta (u-v_l) k}{4}} + \frac{2\pi}{\beta (u-v_l)k} -\frac{2\pi}{\beta (u+v_l)k}  \right],\nonumber     
\end{eqnarray}
which can be simplified for $v_l \ll u$ into 
\begin{eqnarray}\label{eq:attenuation-xxz}
 \alpha_k = \frac{K^2 a^5}{2\pi^2 m v_l u^5 \beta^3} \left(\frac{\partial J^z}{\partial a} -\frac{J^z}{J} \frac{\partial J}{\partial a}\right)^2 (\beta u k)^2 \left[1-\frac{(\beta u k/4)^2}{\sinh(\beta u k/4)^2}\right],   
\end{eqnarray}
yielding the scaling law $\alpha_k = T^3 f(u k/T)$. For $u k \gg k_B T$, $\alpha_k \sim T k^2$ while for $u k \ll T$, $\alpha_k \sim k^4/T$. A maximum of attenuation is obtained for $T \simeq 0.131 \times u k$. The attenuation is represented on Fig.~\ref{fig:attenuation-T}.

\begin{figure}[b]
      \centering
      \includegraphics[width=\linewidth]{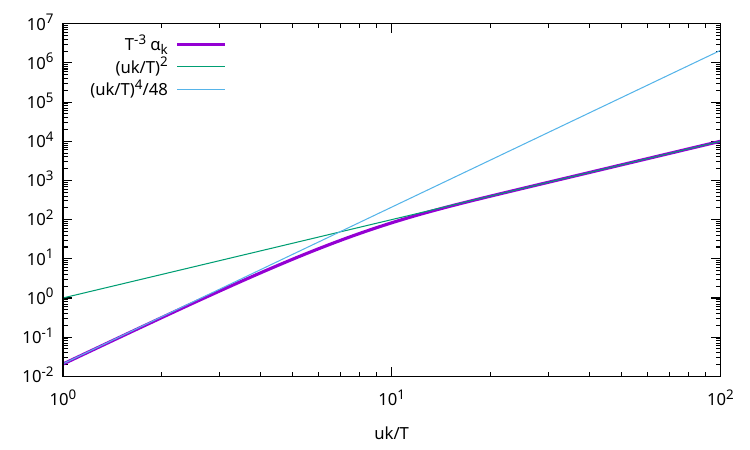}
      \caption{Crossover behavior of the scaling function $T^{-3}\alpha_k=f(uk/T)$ as a function of $uk/T$ for low magnetic field. When the wavelength $2\pi/k$ of the ultrasound exceeds the thermal length $u/T$, the attenuation $\alpha_k \sim k^4/T $, whereas when the wavelength is shorted than $u/T$ the attenuation goes as $\alpha_k \sim T k^2$.  }
      \label{fig:cross-att}
  \end{figure}

The scaling function $f$ in Eq.~(\ref{eq:attenuation}) is universal, in the sense that its functional form does not depend on the Tomonaga-Luttinger exponent. It is represented on Fig.~\ref{fig:cross-att}.  
 In the case of the Heisenberg chain, we have\cite{giamarchi_book_1d} 
 \begin{eqnarray}
 \frac{\partial J}{\partial a} \vec{S}_j\cdot \vec{S}_{j+1} = \frac{\partial J}{J \partial a} \left[ \frac{\pi u}{4} \Pi^2 + \frac{u}{\pi} (\partial_x \phi)^2 + g \frac{\Pi^2}4 - \frac{2g}{\pi^2}(\partial_x \phi)^2 \right. \nonumber\\ 
 \left. -\frac{2g}{(2\pi a)^2} \cos 4\phi \right],     
 \end{eqnarray}
 so the term proportional to $g$ contributes a term $\partial_x \varphi_R \partial_x \varphi_L$ and a term $\cos 4\phi$ having both correlation functions of the form Eq.~(\ref{eq:correlator-marginal}) up to logarithmic corrections\cite{giamarchi_logs}. So the behavior of the attenuation coefficient derived for the XXZ chain is expected to be the same in the case of the Heisenberg chain
  albeit with a different prefactor $(\partial_a J/J)^2$. In the case of the XY chain however, operators $\partial_x \phi_R \partial_x \phi_L$ or $\cos 4\phi$ are absent from the Hamiltonian, and linearizing the spectrum for the fermions yields a vanishing attenuation. 
  
\begin{figure}
    \centering
    \includegraphics[width=\linewidth]{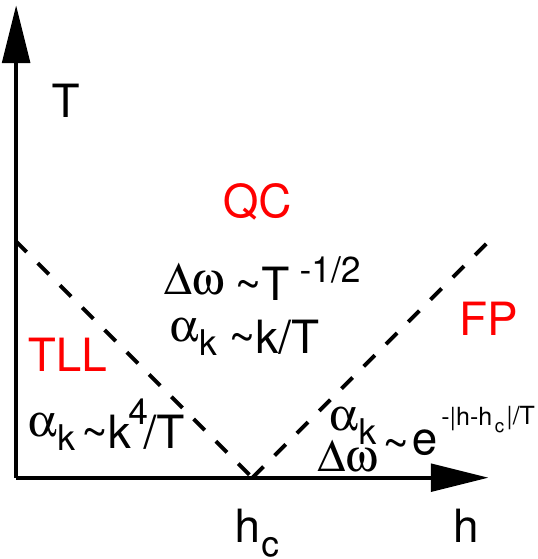}
    \caption{Fan diagram in the vicinity of saturation. In the fully polarized regime (FP), both $\Delta\omega_k$ and $\alpha_k$ are suppressed by the gap $|h+h_c|$. In the Tomonaga Luttinger regime (TLL), $\alpha_k \sim k^4/T$ (under the assumption $uk \ll T$) and $\Delta\omega_k/\omega_k = O(1)$. In the quantum critical regime (QC), $\Delta\omega_k \sim T^{-1/2}$ and $\alpha_k \sim k/T$. The dashed lines are the crossover lines, $T\sim |h-h_c|$. }
    \label{fig:fan-diagram}
\end{figure}

  So far, we have considered only the attenuation in the absence of magnetic field. 
  When we apply a weak magnetic field, it can be absorbed by shifting $\partial_x \phi$ by a constant. This does not modify the correlation function, Eq.~(\ref{eq:correlator-marginal}), so the behavior of the attenuation coefficient remains given by Eq.~(\ref{eq:attenuation-xxz}). With larger fields, the Tomonaga-Luttinger exponent and the velocity become field dependent\cite{haldane_xxzchain} but Eq.~(\ref{eq:attenuation-xxz}) remains valid until the vicinity of saturation\cite{chitra_spinchains_field} where the velocity of spin excitations vanishes as $u(h)\sim |h-h_c|^{1/2}$. The condition $u\gg v_l$ breaks down, and the contributions from $(\partial_x \varphi_{R/L})^2$ that we have neglected can formally become divergent at $u=v_l$. Such divergence, however, is only an artifact of the linearization of the dispersion. Near saturation, the correct low energy field theory is free fermions with a quadratic dispersion\cite{sachdev_qaf_magfield,maeda_universal_2007,he_2017,breunig_quantum_2017}. The resulting  behavior is most easily understood in the case of the XY model. 
In the XY model, using the Jordan-Wigner representation, we have\cite{caprani_absorption_2023} 
\begin{eqnarray}
&&\alpha_k=\frac{2\pi \sin^2(ka/2)}{m v_l a \omega_k} \left(\frac{\partial J}{\partial a} \right)^2 \int_{-\frac \pi a}^{\frac \pi a} dk_1 [n_F(J \cos (k+k_1)a -h) \nonumber \\
&&-n_F (J \cos k_1 a -h) ] \delta(\omega_k + J \cos k_1 a -J \cos (k_1+k) a).          
\end{eqnarray}
When $h\simeq -J$, since $J \cos (\pi +\delta k a)  -h \simeq J (\delta k a)^2 -(J+h)$ for $\delta k a \ll 1$ we obtain the approximation
\begin{widetext}
\begin{equation}
\alpha_k = \frac{2 \pi \sin^2 (ka/2)}{m J v_l^2 k^2 a^2} \left(\frac{\partial J}{\partial a} \right)^2   \frac {\sinh (\beta v_l k/2)} {\cosh(\beta v_l k/2) + \cosh(\beta (J+h) -\beta J a^2 k^2/8 -\beta v_l^2/(2 J a^2)) }.     
\end{equation}
\end{widetext}
Away from saturation, with $|J+h| \gg T$, $\alpha_k \sim e^{-|J+h|/T}$, instead of $\alpha_k=0$ obtained with linearization. But near saturation, with $|J+h| \ll T$, $\alpha_k \sim k/T$, and sound attenuation is strongly enhanced. Because the free fermions with quadratic dispersion fixed point also describes the saturation in  spin chains and the vicinity of plateau transitions transition\cite{chitra_spinchains_field,oshikawa_plateaus}, a similar enhancement of attenuation is expected in these limits. Combining the results derived in the Luttinger liquid phase for $\alpha_k$ with the results close to saturation, we can draw a fan diagram representing the different regimes in the vicinity of the quantum critical point on Fig.~\ref{fig:fan-diagram}.  

\section{Conclusion}
We have considered attenuation and velocity shift of longitudinal phonons in an antiferromagnetic spin-1/2 XXZ chain. We have related the velocity shift to derivatives of the free energy of the spin chain. In the Tomonaga-Luttinger liquid regime, this implies that the shift behaves as an affine function of the temperature squared at low temperature. Near saturation, instead, the shift diverges as the inverse of the square root of the temperature. In a fully saturated chain, the shift is strongly suppressed by the Boltzmann factor. Such behavior can be generally expected in other one-dimensional systems showing a magnetization plateau\cite{oshikawa_plateaus,cabra_magnetization_plateaus} near the plateau end as a consequence of the presence of a $z=2$ quantum critical point.\cite{sachdev_qaf_magfield,chitra_spinchains_field,he_2017,breunig_quantum_2017,zheludev_2020} For temperatures of the order of the exchange constant, the shift can be calculated by solving a non-linear integral equation.\cite{kluemper_thermo_xxx} For the attenuation, we have found that in a Tomonaga-Luttinger liquid, it would behave as $T k^2$ when the wavevector $k a \gg T/J $ and $k^4/T$ when $ka \ll T/J$ where $a$ is the lattice spacing, and $J$ the exchange interaction. In the vicinity of saturation, the attenuation is expected to behave as $k/T$. All those results are universal in the sense that no anomalous exponent depending on the Tomonaga-Luttinger exponent are predicted. 
The results we have derived in the Tomonaga-Luttinger liquid phase have broader applicability than the XXZ spin-1/2 chain in the easy plane regime. Tomonaga-Luttinger liquids can be realised under high enough field\cite{zapf_2014} in XXZ spin-1/2 chain in the easy axis regime \cite{klanjsek_2015,faure_2019}, in Haldane gapped chains\cite{zheludev_ndmap_2002,chiatti_character_2008,bera_field_2013,li_haldane_2025}, spin ladders\cite{Klanjsek-PRL-2008,hong:2010}, frustrated\cite{chen_zigzag_2024} and dimerized spin-1/2 chains\cite{yoshida_f5pnn_2005,stone_cunitrate_2014,willenberg_2015,islam_2024} and the transition under applied magnetic field between the gapped phase and the  Tomonaga-Luttinger liquid is in the same universality class as the transition near saturation in the XXZ spin-1/2 chain\cite{japaridze_cic_transition,pokrovsky_talapov_prl,sachdev_qaf_magfield,he_2017,breunig_quantum_2017,chitra_spinchains_field}. A different universality class can be realized\cite{pfeuty_transverse_ising,dutta_book_2015} with Ising spin chains in a transverse field\cite{matsuura_conb2o6_2020}. In some ladders\cite{clemancey06_cuhpcl}, staggered Dzyaloshinskii-Moriya interactions\cite{dzyalo_interaction,moriya60} can turn the critical point into a crossover. These cases could be analyzed by extending the approach of the present paper\cite{orignac07_staggered_dm}. Our study has focused on decoupled chains, with longitudinal acoustic phonons propagating along the chains. At high enough temperature, the propagation of transverse phonons along the chains  
or longitudinal phonons in the directions orthogonal to the chains 
could be considered using perturbation theory in interchain coupling. 
In such case, anomalous exponents depending on Tomonaga-Luttinger parameter can be expected. 
At lower temperatures, the Tomonaga-Luttinger liquid can become unstable leading to either antiferromagnetic ordering\cite{giamarchi_coupled_ladders,irkhin00_chain_mf,poirier_2002,Klanjsek-PRL-2008} or  spin-Peiers ordering via  optical phonon softening\cite{pincus_spin-peierls,pytte_sp,cross_spinpeierls,inagaki_spinpeierls,orignac04_tbamf,citro_2005}. In classical thermodynamics, anomalies of the sound velocity near phase transitions are well known\cite{pawlak_2009}. Experiments in deuterated $(\mathrm{TMTTF})_2\mathrm{PF}_6$ have revealed the presence of a dip\cite{poirier_2012} in the relative velocity shift at the spin-Peierls transition temperature. Similarly, anomalies\cite{poirier_2002,yamaguchi_2011} of the velocity shift have been observed in $\mathrm{BaCu_2Si_2O_7}$ and $\mathrm{BaCo_2V_2O_8}$ near the N\'eel transition.  To capture such effects, in particular the pretransitional fluctuations, non-perturbative treatments of interchain interactions are necessary\cite{inagaki_spinpeierls,schulz_coupled_spinchains,giamarchi_coupled_ladders,irkhin00_chain_mf,bocquet02_chain_rpa,dupont2018}. These questions are left for future studies.

\begin{acknowledgments}
We thank S. Galeski for discussions and for sharing some unpublished experimental data. We thank S. Zherlitsyn, M. Brando, A. Haubsburg, M. Vergniory for discussions. 
\end{acknowledgments}

\appendix

\section{Response and relaxation function}\label{app:response} 
If we consider the relaxation function and insert a resolution of the
identity under the trace,
\begin{eqnarray}
  \label{eq:relax}
 && (A(t),A^\dagger(0))=\int_0^\beta \frac{d\lambda} Z \sum_{n,m} e^{-\beta E_m} e^{(\lambda+it) (E_n-E_m)} |\langle n|A|m\rangle|^2 \nonumber \\
  &&= \frac 1 Z \sum_{n,m} \frac{e^{-\beta E_m}-e^{-\beta E_n}}{E_n-E_m} e^{it (E_n-E_m)}  |\langle n|A|m\rangle|^2.  
\end{eqnarray}
Taking a Laplace transform with $z=\omega + i 0_+$,
\begin{eqnarray}
 \int_0^{+\infty} dt e^{iz t}  (A(t),A^\dagger(0))= \frac i {Z} \sum_{n,m}  \frac{e^{-\beta E_m}-e^{-\beta E_n}}{E_n-E_m} \frac{ |\langle n|A|m\rangle|^2}{z+E_n-E_m}, 
\end{eqnarray}
and the real part is
\begin{eqnarray}
  &&\mathrm{Re} \int_0^{+\infty} dt e^{iz t}  (A(t),A^\dagger(0)) \nonumber \\
  &&= \frac{\pi}{Z\omega} \sum_{n,m} (e^{-\beta E_n} - e^{-\beta E_m})  |\langle n|A|m\rangle|^2\delta(\omega + E_n -E_m).  
\end{eqnarray}

If we consider the response function
\begin{eqnarray}
  \chi_A(t) = i\theta(t) \langle [A(t),A^\dagger(0)]\rangle,  
\end{eqnarray}
we have
\begin{eqnarray}
  \mathrm{Im} \int_0^{+\infty}  dt  e^{izt} \chi_A(t) &=& \frac{\pi}{Z} \sum_{n,m} (e^{-\beta E_n} - e^{-\beta E_m})  |\langle n|A|m\rangle|^2\nonumber \\ 
  &&\times \delta(\omega + E_n -E_m)\nonumber \\
  &=& \omega   \mathrm{Re} \int_0^{+\infty} dt e^{iz t}  (A(t),A^\dagger(0)). 
\end{eqnarray}
Using $(b_{k},b_{k})=1/\omega_{k}^0$, and the definition\cite{tachiki_1974}, we can therefore rewrite the sound attenuation
\begin{eqnarray}\label{attenuation-response} 
  \alpha_{k} =\lim_{\delta \to 0_+}  \mathrm{Im }\int_0^{\infty} \frac{dt}{v_l} \langle [f_{k}(t), f^\dagger_{k}(0)]\rangle e^{-i\omega_k t} e^{-\delta t}.  
\end{eqnarray}
\section{Velocity shift}\label{app:vshift}
The Matsubara Green's function of phonons is\cite{abrikosov_book,mahan_book}
\begin{eqnarray}
\mathcal{D}(k,\tau)&=&-\frac{\omega_k} 2 \langle T_\tau (b_k + b^\dagger_{-k}(\tau) (b_{-k}+b^\dagger_k)(0)\rangle, \\ 
&=&\frac 1 {\beta} \sum_{i\omega_n} \mathcal{D}^{0)}(k,i\omega_n) e^{-i\omega_n \tau}. 
\end{eqnarray}
In the absence of spin-phonon interaction, the Matsubara Green's function of the phonons in frequency representation is\cite{abrikosov_book,mahan_book} 
\begin{equation}\label{eq:phonon-green-bare}
\mathcal{D}^{(0)}(k,i\omega_n)=-\frac{\omega_k^2}{\omega_n^2+\omega_k^2}.    
\end{equation}
In the presence of spin-phonon interactions Eq.~(\ref{eq:u1-decoupled})--(\ref{eq:u2-decoupled}), the Green's function becomes 
\begin{equation}
\mathcal{D}(k,i\omega_n)=\frac{1}{(\mathcal{D}^{(0)})^{-1}(k,\omega_n) -P(k,\omega_n) },     
\end{equation}
where $P(k,\omega_n)$ is the phonon self-energy. After analytic continuation, the poles of the Green's function are given by 
\begin{equation}
1-\frac{\omega^2}{\omega_k}^2 -P(k, i\omega_n \to \omega) = 0.     
\end{equation}
Perturbatively, we set $\omega=\omega_k+(\Delta \omega)_k + \frac{i}{\tau_k}$, 
and we find at lowest order
\begin{eqnarray}
(\Delta \omega)_k = -\frac{\omega_k} 2 \mathrm{Re} P(k,\omega_k+iO_+), \\ 
\frac{1}{\tau_k} = -\frac{\omega_k} 2 \mathrm{Im} P(k,\omega_k+iO_+). \\ 
\end{eqnarray}
Analogous formulas are used in the vicinity of phase transitions in classical systems.\cite{pawlak_2009}
Applying the usual rules of diagrammatic perturbation theory\cite{abrikosov_book,mahan_book}, the self energy is 
\begin{equation}
P(k,i\omega_n)=\frac{2}{Nm\omega_k^2}\left[\langle U_{{kk}}^{(2)} \rangle -\frac 1 2 \int_0^\beta \langle T_\tau U_k^{(1)}(\tau) U_{-k}^{(1)}(0)\rangle e^{i\omega_n \tau} \right],     
\end{equation}
to first order in $H_2$ and second order in $H_1$. After analytic continuation, we recover Eqs.~(\ref{eq:shift2-tachiki})--(\ref{eq:shift1-quantum}), giving the shift in pulsation. 
If we turn to $\tau_k$, by comparing with Eq.~(\ref{eq:att-matsubara}) that $\alpha_k=1/(v_l \tau_k)$. 

\section{Derivatives of the free energy in the Heisenberg spin chain}
\label{section:NLE}
\subsection{Ground state}
The ground state energy  $e_0(h,J)$ of the Heisenberg spin chain and is obtained by solving a linear integral equation\cite{kluemper_thermo_xxx,cabra_magnetization_plateaus},
\begin{eqnarray}
\rho(x)=\frac 1 {\pi(x^2+1)} -\int_{-Q}^{Q} \frac{dy}{2\pi} \frac{4}{4+(x-y)^2} \rho(y),    
\end{eqnarray}
with magnetization given by 
\begin{equation}
 m=\frac 1 2 -\int_{-Q}^Q dx \rho(x).     
\end{equation}
The magnetic field is obtained from the auxiliary integral equation  
\begin{eqnarray}
\xi(x)=1-\int_{-Q}^{Q} \frac{dy}{2\pi} \frac{4}{4+(x-y)^2} \xi(y), 
\end{eqnarray}
and $h=2\pi J \rho(Q)/\xi(Q)$. Inverting the relation gives $Q$ as a function of $h/J$.  The energy is
\begin{equation}
e_0=-2J \int_{-Q}^{Q}\frac{\rho(x) dx}{x^2+1}-h m,      
\end{equation}
so $e_0(J,h)=J\epsilon_0(h/J)$. We have the relations 
\begin{eqnarray}
\frac{\partial e_0}{\partial J}=\frac{\epsilon_0}{J} -\frac{h}{J}\frac{\partial\epsilon_0}{\partial h},\\
\frac{\partial^2 e_0}{\partial J^2}=-\frac{h^2} {J^2} \frac{\partial^2\epsilon_0}{\partial h^2},\\
\end{eqnarray}
that correspond to the zero temperature limit of~(\ref{eq:thermo-field-deriv-1})--(\ref{eq:thermo-field-deriv-2}). 
\subsection{Finite temperature}
To calculate the free energy of the Heisenberg chain in a magnetic field, we must solve the coupled integral equations\cite{kluemper_thermo_xxx} that are analytically derived by exploiting the Bethe Ansatz to diagonalize the Quantum Transfer Matrix (QTM), thereby reducing an infinite set of thermodynamic equations into just two coupled equations governed by the QTM's dominant eigenvalue:
\begin{widetext}
\begin{eqnarray}\label{eq:kluemper-nlie}
\ln a(x) = -\frac{u \beta}{\cosh x } + \frac{\beta h } 2 + \int_{-\infty}^\infty dy \left\{k(x-y) \ln[1+a(y)] + k(x-y-i\pi + i\epsilon) \ln[1+\bar{a}(y)] \right\}, \\
\ln \bar{a}(x) = -\frac{u \beta}{\cosh x } - \frac{\beta h } 2 + \int_{-\infty}^\infty dy \left\{k(x-y) \ln[1+\bar{a}(y)] + k(x-y+i\pi - i\epsilon) \ln[1+a(y)]  \right\}, 
\end{eqnarray}
\end{widetext}
where $u=\pi J/2$ and the kernel 
\begin{equation}
    k(x)= \int_{-\infty}^\infty \frac{dk}{4\pi} \frac{e^{-\frac \pi 2 |k|}}{\cosh \left(\frac {\pi k } 2\right)} e^{ikx}.  
\end{equation}
the free energy is given by the integral 
\begin{equation}
F= e_0(J,h=0)  -\frac {T}{2\pi} \int \frac{dx}{\cosh x} \log [(1+a(x))(1+\bar{a}(x)],       
\end{equation}

The derivatives of the free energy are obtained from the integral equations by partial differentiation with respect to $J$ of Eq.~(\ref{eq:kluemper-nlie}). We have the equations for the first derivatives
\begin{widetext}
\begin{eqnarray}
\frac 1 {a(x)} \frac{\partial a(x)}{ \partial J} =   -\frac{\pi \beta}{2 \cosh x } + \int_{-\infty}^\infty dy \left\{k(x-y) \frac 1 {1+a(y)} \frac{\partial a(y)}{ \partial J} +  k(x-y-i\pi + i\epsilon) \frac 1 {1+\bar{a}(y)} ] \frac{\partial \bar{a}(y)}{ \partial J} \right\}, \\ 
\frac 1 {\bar{a}(x)} \frac{\partial \bar{a}(x)}{ \partial J} =   -\frac{\pi \beta}{2 \cosh x } + \int_{-\infty}^\infty dy \left\{k(x-y) \frac 1 {1+\bar{a}(y)}  \frac{\partial \bar{a}(y)}{ \partial J}  +  k(x-y+i\pi - i\epsilon) \frac 1 {1+a(y)} \frac{\partial a(y)}{ \partial J} \right\}, \\ 
\end{eqnarray}
\end{widetext}
that allow us to obtain $\partial F/\partial J$.
For the second derivatives, 
\begin{widetext}
\begin{eqnarray}
\frac 1 a \frac{\partial^2 a}{\partial^2 J} = \left(\frac 1 a \frac{\partial a}{\partial J} \right)^2 + \int_{-\infty}^\infty dy \left\{k(x-y) \frac{\partial}{\partial J} \left[\frac 1 {1+a(y)} \frac{\partial a(y)}{ \partial J} \right]  +  k(x-y-i\pi + i\epsilon) \frac{\partial}{\partial J} \left[\frac 1 {1+\bar{a}(y)} ] \frac{\partial \bar{a}(y)}{ \partial J} \right] \right\}, \\  
\frac 1 a \frac{\partial^2 a}{\partial^2 J} = \left(\frac 1 a \frac{\partial a}{\partial J} \right)^2 + \int_{-\infty}^\infty dy \left\{k(x-y) \frac{\partial}{\partial J} \left[\frac 1 {1+\bar{a}(y)} \frac{\partial \bar{a}(y)}{ \partial J} \right]  +  k(x-y+i\pi - i\epsilon) \frac{\partial}{\partial J} \left[\frac 1 {1+a(y)} ] \frac{\partial a(y)}{ \partial J} \right] \right\}, \\ 
\end{eqnarray}
\end{widetext}
allow us to find $\partial^2 a/\partial^2 J$ and $\partial^2 \bar a /\partial^2 J$ knowing $\partial a/\partial J$ and $\partial \bar{a}/\partial J$ and calculate $\partial^2 F/\partial^2 J$.
To solve these integral equations by relaxation methods\cite{press92_nr_nystrom}, we wrote a Fortran 90 program. The equations were discretized on a a finite interval. The convolutions in Eq.~(\ref{eq:kluemper-nlie}) were performed by fast Fourier transforms using the CFFT routines of the SLATEC mathematical library.\cite{fong_guide_1993} Once the functions $a(x),\bar{a}(x)$ are obtained, the successive integral equations for their derivatives are solved, allowing to calculate the free energy and its derivatives. 


\begin{thebibliography}{87}
\expandafter\ifx\csname natexlab\endcsname\relax\def\natexlab#1{#1}\fi
\expandafter\ifx\csname bibnamefont\endcsname\relax
  \def\bibnamefont#1{#1}\fi
\expandafter\ifx\csname bibfnamefont\endcsname\relax
  \def\bibfnamefont#1{#1}\fi
\expandafter\ifx\csname citenamefont\endcsname\relax
  \def\citenamefont#1{#1}\fi
\expandafter\ifx\csname url\endcsname\relax
  \def\url#1{\texttt{#1}}\fi
\expandafter\ifx\csname urlprefix\endcsname\relax\def\urlprefix{URL }\fi
\providecommand{\bibinfo}[2]{#2}
\providecommand{\eprint}[2][]{\url{#2}}

\bibitem[{\citenamefont{Tani and Mori}(1968)}]{tani_1968}
\bibinfo{author}{\bibfnamefont{K.}~\bibnamefont{Tani}} \bibnamefont{and}
  \bibinfo{author}{\bibfnamefont{H.}~\bibnamefont{Mori}},
  \bibinfo{journal}{Prog. Theor. Phys.} \textbf{\bibinfo{volume}{34}},
  \bibinfo{pages}{876} (\bibinfo{year}{1968}).

\bibitem[{\citenamefont{Kawasaki and Ikushima}(1970)}]{kawasaki_1970}
\bibinfo{author}{\bibfnamefont{K.}~\bibnamefont{Kawasaki}} \bibnamefont{and}
  \bibinfo{author}{\bibfnamefont{A.}~\bibnamefont{Ikushima}},
  \bibinfo{journal}{Phys. Rev. B} \textbf{\bibinfo{volume}{1}},
  \bibinfo{pages}{3143} (\bibinfo{year}{1970}).

\bibitem[{\citenamefont{Tachiki and Maekawa}(1974)}]{tachiki_1974}
\bibinfo{author}{\bibfnamefont{M.}~\bibnamefont{Tachiki}} \bibnamefont{and}
  \bibinfo{author}{\bibfnamefont{S.}~\bibnamefont{Maekawa}},
  \bibinfo{journal}{Prog. Theor. Phys.} \textbf{\bibinfo{volume}{51}},
  \bibinfo{pages}{1} (\bibinfo{year}{1974}).

\bibitem[{\citenamefont{Pawlak}(2009)}]{pawlak_2009}
\bibinfo{author}{\bibfnamefont{A.}~\bibnamefont{Pawlak}}, in
  \emph{\bibinfo{booktitle}{Horizons in World Physics}}, edited by
  \bibinfo{editor}{\bibfnamefont{L.}~\bibnamefont{Pedroza}} \bibnamefont{and}
  \bibinfo{editor}{\bibfnamefont{M.}~\bibnamefont{Everett}}
  (\bibinfo{publisher}{Nova Science Publishers}, \bibinfo{address}{Hauppauge,
  NY}, \bibinfo{year}{2009}), vol. \bibinfo{volume}{268},
  p.~\bibinfo{pages}{69}.

\bibitem[{\citenamefont{Poirier et~al.}(2002)\citenamefont{Poirier, Castonguay,
  Revcolevschi, and Dhalenne}}]{poirier_2002}
\bibinfo{author}{\bibfnamefont{M.}~\bibnamefont{Poirier}},
  \bibinfo{author}{\bibfnamefont{M.}~\bibnamefont{Castonguay}},
  \bibinfo{author}{\bibfnamefont{A.}~\bibnamefont{Revcolevschi}},
  \bibnamefont{and} \bibinfo{author}{\bibfnamefont{G.}~\bibnamefont{Dhalenne}},
  \bibinfo{journal}{Phys. Rev. B} \textbf{\bibinfo{volume}{66}},
  \bibinfo{pages}{054402} (\bibinfo{year}{2002}).

\bibitem[{\citenamefont{Wolf et~al.}(2004)\citenamefont{Wolf, Zherlitsyn,
  Lüthi, Harrison, Löw, Pashchenko, Lang, Margraf, Lerner, Dahlmann
  et~al.}}]{wolf_2004}
\bibinfo{author}{\bibfnamefont{B.}~\bibnamefont{Wolf}},
  \bibinfo{author}{\bibfnamefont{S.}~\bibnamefont{Zherlitsyn}},
  \bibinfo{author}{\bibfnamefont{B.}~\bibnamefont{Lüthi}},
  \bibinfo{author}{\bibfnamefont{N.}~\bibnamefont{Harrison}},
  \bibinfo{author}{\bibfnamefont{U.}~\bibnamefont{Löw}},
  \bibinfo{author}{\bibfnamefont{V.}~\bibnamefont{Pashchenko}},
  \bibinfo{author}{\bibfnamefont{M.}~\bibnamefont{Lang}},
  \bibinfo{author}{\bibfnamefont{G.}~\bibnamefont{Margraf}},
  \bibinfo{author}{\bibfnamefont{H.-W.} \bibnamefont{Lerner}},
  \bibinfo{author}{\bibfnamefont{E.}~\bibnamefont{Dahlmann}},
  \bibnamefont{et~al.}, \bibinfo{journal}{Phys. Rev. B}
  \textbf{\bibinfo{volume}{69}}, \bibinfo{pages}{092403}
  (\bibinfo{year}{2004}).

\bibitem[{\citenamefont{Chiatti et~al.}(2008)\citenamefont{Chiatti, Sytcheva,
  Wosnitza, Zherlitsyn, Zvyagin, Zapf, Jaime, and
  Paduan-Filho}}]{chiatti_character_2008}
\bibinfo{author}{\bibfnamefont{O.}~\bibnamefont{Chiatti}},
  \bibinfo{author}{\bibfnamefont{A.}~\bibnamefont{Sytcheva}},
  \bibinfo{author}{\bibfnamefont{J.}~\bibnamefont{Wosnitza}},
  \bibinfo{author}{\bibfnamefont{S.}~\bibnamefont{Zherlitsyn}},
  \bibinfo{author}{\bibfnamefont{A.~A.} \bibnamefont{Zvyagin}},
  \bibinfo{author}{\bibfnamefont{V.~S.} \bibnamefont{Zapf}},
  \bibinfo{author}{\bibfnamefont{M.}~\bibnamefont{Jaime}}, \bibnamefont{and}
  \bibinfo{author}{\bibfnamefont{A.}~\bibnamefont{Paduan-Filho}},
  \bibinfo{journal}{Phys. Rev. B} \textbf{\bibinfo{volume}{78}},
  \bibinfo{pages}{094406} (\bibinfo{year}{2008}).

\bibitem[{\citenamefont{Yamaguchi et~al.}(2011)\citenamefont{Yamaguchi, Yasin,
  Zherlitsyn, Omura, Kimura, Yoshii, Okunishi, He, Taniyama, Itoh
  et~al.}}]{yamaguchi_2011}
\bibinfo{author}{\bibfnamefont{H.}~\bibnamefont{Yamaguchi}},
  \bibinfo{author}{\bibfnamefont{S.}~\bibnamefont{Yasin}},
  \bibinfo{author}{\bibfnamefont{S.}~\bibnamefont{Zherlitsyn}},
  \bibinfo{author}{\bibfnamefont{K.}~\bibnamefont{Omura}},
  \bibinfo{author}{\bibfnamefont{S.}~\bibnamefont{Kimura}},
  \bibinfo{author}{\bibfnamefont{S.}~\bibnamefont{Yoshii}},
  \bibinfo{author}{\bibfnamefont{K.}~\bibnamefont{Okunishi}},
  \bibinfo{author}{\bibfnamefont{Z.}~\bibnamefont{He}},
  \bibinfo{author}{\bibfnamefont{T.}~\bibnamefont{Taniyama}},
  \bibinfo{author}{\bibfnamefont{M.}~\bibnamefont{Itoh}}, \bibnamefont{et~al.},
  \bibinfo{journal}{J. Phys. Soc. Jpn.} \textbf{\bibinfo{volume}{80}},
  \bibinfo{pages}{033701} (\bibinfo{year}{2011}). 
  

\bibitem[{\citenamefont{Poirier et~al.}(2012)\citenamefont{Poirier, Langlois,
  Bourbonnais, Foury-Leylekian, Moradpour, and Pouget}}]{poirier_2012}
\bibinfo{author}{\bibfnamefont{M.}~\bibnamefont{Poirier}},
  \bibinfo{author}{\bibfnamefont{A.}~\bibnamefont{Langlois}},
  \bibinfo{author}{\bibfnamefont{C.}~\bibnamefont{Bourbonnais}},
  \bibinfo{author}{\bibfnamefont{P.}~\bibnamefont{Foury-Leylekian}},
  \bibinfo{author}{\bibfnamefont{A.}~\bibnamefont{Moradpour}},
  \bibnamefont{and} \bibinfo{author}{\bibfnamefont{J.-P.}
  \bibnamefont{Pouget}}, \bibinfo{journal}{Phys. Rev. B}
  \textbf{\bibinfo{volume}{86}}, \bibinfo{pages}{085111}
  (\bibinfo{year}{2012}), \bibinfo{note}{arXiv: 1207.6361}. 

\bibitem[{\citenamefont{Sergeicheva et~al.}(2020)\citenamefont{Sergeicheva,
  Sosin, Gorbunov, Zherlitsyn, Gu, and Zaliznyak}}]{sergeicheva_2020}
\bibinfo{author}{\bibfnamefont{E.~G.} \bibnamefont{Sergeicheva}},
  \bibinfo{author}{\bibfnamefont{S.~S.} \bibnamefont{Sosin}},
  \bibinfo{author}{\bibfnamefont{D.~I.} \bibnamefont{Gorbunov}},
  \bibinfo{author}{\bibfnamefont{S.}~\bibnamefont{Zherlitsyn}},
  \bibinfo{author}{\bibfnamefont{G.}~\bibnamefont{Gu}}, \bibnamefont{and}
  \bibinfo{author}{\bibfnamefont{I.}~\bibnamefont{Zaliznyak}},
  \bibinfo{journal}{Phys. Rev. B} \textbf{\bibinfo{volume}{101}},
  \bibinfo{pages}{201107} (\bibinfo{year}{2020}), \bibinfo{note}{arXiv:
  1911.07592}.

\bibitem[{\citenamefont{Povarov et~al.}(2024)\citenamefont{Povarov, Graf,
  Hauspurg, Zherlitsyn, Wosnitza, Sakurai, Ohta, Kimura, Nojiri, Garlea
  et~al.}}]{povarov_2024}
\bibinfo{author}{\bibfnamefont{K.~Y.} \bibnamefont{Povarov}},
  \bibinfo{author}{\bibfnamefont{D.~E.} \bibnamefont{Graf}},
  \bibinfo{author}{\bibfnamefont{A.}~\bibnamefont{Hauspurg}},
  \bibinfo{author}{\bibfnamefont{S.}~\bibnamefont{Zherlitsyn}},
  \bibinfo{author}{\bibfnamefont{J.}~\bibnamefont{Wosnitza}},
  \bibinfo{author}{\bibfnamefont{T.}~\bibnamefont{Sakurai}},
  \bibinfo{author}{\bibfnamefont{H.}~\bibnamefont{Ohta}},
  \bibinfo{author}{\bibfnamefont{S.}~\bibnamefont{Kimura}},
  \bibinfo{author}{\bibfnamefont{H.}~\bibnamefont{Nojiri}},
  \bibinfo{author}{\bibfnamefont{V.~O.} \bibnamefont{Garlea}},
  \bibnamefont{et~al.}, \bibinfo{journal}{Nature Communications}
  \textbf{\bibinfo{volume}{15}}, \bibinfo{pages}{2295} (\bibinfo{year}{2024}),
  \bibinfo{note}{arXiv:2306.15450 [cond-mat.str-el]}.

\bibitem[{\citenamefont{Takahashi}(1999)}]{takahashi_book_bethe}
\bibinfo{author}{\bibfnamefont{M.}~\bibnamefont{Takahashi}},
  \emph{\bibinfo{title}{Thermodynamics of One-Dimensional Solvable Models}}
  (\bibinfo{publisher}{Cambridge University Press},
  \bibinfo{address}{Cambridge}, \bibinfo{year}{1999}).

\bibitem[{\citenamefont{Tsyplyatyev et~al.}(2017)\citenamefont{Tsyplyatyev,
  Kopietz, Tsui, Wolf, Cong, van Well, Ritter, Krellner, Aßmus, and
  Lang}}]{tsyplyatyev_2017}
\bibinfo{author}{\bibfnamefont{O.}~\bibnamefont{Tsyplyatyev}},
  \bibinfo{author}{\bibfnamefont{P.}~\bibnamefont{Kopietz}},
  \bibinfo{author}{\bibfnamefont{Y.}~\bibnamefont{Tsui}},
  \bibinfo{author}{\bibfnamefont{B.}~\bibnamefont{Wolf}},
  \bibinfo{author}{\bibfnamefont{P.~T.} \bibnamefont{Cong}},
  \bibinfo{author}{\bibfnamefont{N.}~\bibnamefont{van Well}},
  \bibinfo{author}{\bibfnamefont{F.}~\bibnamefont{Ritter}},
  \bibinfo{author}{\bibfnamefont{C.}~\bibnamefont{Krellner}},
  \bibinfo{author}{\bibfnamefont{W.}~\bibnamefont{Aßmus}}, \bibnamefont{and}
  \bibinfo{author}{\bibfnamefont{M.}~\bibnamefont{Lang}},
  \bibinfo{journal}{Phys. Rev. B} \textbf{\bibinfo{volume}{95}},
  \bibinfo{pages}{045120} (\bibinfo{year}{2017}). 

\bibitem[{\citenamefont{Citro et~al.}(2005{\natexlab{a}})\citenamefont{Citro,
  Orignac, and Giamarchi}}]{citro04_spinpeierls}
\bibinfo{author}{\bibfnamefont{R.}~\bibnamefont{Citro}},
  \bibinfo{author}{\bibfnamefont{E.}~\bibnamefont{Orignac}}, \bibnamefont{and}
  \bibinfo{author}{\bibfnamefont{T.}~\bibnamefont{Giamarchi}},
  \bibinfo{journal}{Phys. Rev. B} \textbf{\bibinfo{volume}{72}},
  \bibinfo{pages}{024434} (\bibinfo{year}{2005}{\natexlab{a}}),
  \eprint{cond-mat/0411256}.

\bibitem[{\citenamefont{Giamarchi}(2004)}]{giamarchi_book_1d}
\bibinfo{author}{\bibfnamefont{T.}~\bibnamefont{Giamarchi}},
  \emph{\bibinfo{title}{Quantum Physics in One Dimension}}
  (\bibinfo{publisher}{Oxford University Press}, \bibinfo{address}{Oxford},
  \bibinfo{year}{2004}).

\bibitem[{\citenamefont{Destri and de~Vega}(1992)}]{destri92_nostrings}
\bibinfo{author}{\bibfnamefont{C.}~\bibnamefont{Destri}} \bibnamefont{and}
  \bibinfo{author}{\bibfnamefont{H.}~\bibnamefont{de~Vega}},
  \bibinfo{journal}{Phys. Rev. Lett.} \textbf{\bibinfo{volume}{69}},
  \bibinfo{pages}{2313} (\bibinfo{year}{1992}).

\bibitem[{\citenamefont{Destri and de~Vega}(1995)}]{destri95_tba_inteq}
\bibinfo{author}{\bibfnamefont{C.}~\bibnamefont{Destri}} \bibnamefont{and}
  \bibinfo{author}{\bibfnamefont{H.}~\bibnamefont{de~Vega}},
  \bibinfo{journal}{Nucl. Phys. B} \textbf{\bibinfo{volume}{438}},
  \bibinfo{pages}{413} (\bibinfo{year}{1995}).

\bibitem[{\citenamefont{{Kl{\"u}mper}}(1998)}]{kluemper_heisenberg_thermo}
\bibinfo{author}{\bibfnamefont{A.}~\bibnamefont{{Kl{\"u}mper}}},
  \bibinfo{journal}{Eur. Phys. J. B} \textbf{\bibinfo{volume}{5}},
  \bibinfo{pages}{677} (\bibinfo{year}{1998}).

\bibitem[{\citenamefont{{Kl{\"u}mper} and
  Johnston}(2000)}]{kluemper_thermo_xxx}
\bibinfo{author}{\bibfnamefont{A.}~\bibnamefont{{Kl{\"u}mper}}}
  \bibnamefont{and} \bibinfo{author}{\bibfnamefont{D.~C.}
  \bibnamefont{Johnston}}, \bibinfo{journal}{Phys. Rev. Lett.}
  \textbf{\bibinfo{volume}{84}}, \bibinfo{pages}{4701} (\bibinfo{year}{2000}).

\bibitem[{\citenamefont{Kluemper and Sakai}(2002)}]{kluemper_xxz}
\bibinfo{author}{\bibfnamefont{A.}~\bibnamefont{Kluemper}} \bibnamefont{and}
  \bibinfo{author}{\bibfnamefont{K.}~\bibnamefont{Sakai}}, \bibinfo{journal}{J.
  Phys. A} \textbf{\bibinfo{volume}{35}}, \bibinfo{pages}{2173}
  (\bibinfo{year}{2002}).

\bibitem[{\citenamefont{Bouchoule et~al.}(2025)\citenamefont{Bouchoule, Citro,
  Duty, Giamarchi, Hulet, Klanj{\v{s}}ek, Orignac, and Weber}}]{Bouchoule2025}
\bibinfo{author}{\bibfnamefont{I.}~\bibnamefont{Bouchoule}},
  \bibinfo{author}{\bibfnamefont{R.}~\bibnamefont{Citro}},
  \bibinfo{author}{\bibfnamefont{T.}~\bibnamefont{Duty}},
  \bibinfo{author}{\bibfnamefont{T.}~\bibnamefont{Giamarchi}},
  \bibinfo{author}{\bibfnamefont{R.~G.} \bibnamefont{Hulet}},
  \bibinfo{author}{\bibfnamefont{M.}~\bibnamefont{Klanj{\v{s}}ek}},
  \bibinfo{author}{\bibfnamefont{E.}~\bibnamefont{Orignac}}, \bibnamefont{and}
  \bibinfo{author}{\bibfnamefont{B.}~\bibnamefont{Weber}},
  \bibinfo{journal}{Nature Reviews Physics} \textbf{\bibinfo{volume}{7}},
  \bibinfo{pages}{565} (\bibinfo{year}{2025}). 

\bibitem[{\citenamefont{Landau and
  Lifshitz}(1959{\natexlab{a}})}]{landau_statmech}
\bibinfo{author}{\bibfnamefont{L.~D.} \bibnamefont{Landau}} \bibnamefont{and}
  \bibinfo{author}{\bibfnamefont{E.~M.} \bibnamefont{Lifshitz}},
  \emph{\bibinfo{title}{Statistical Physics}} (\bibinfo{publisher}{Pergamon
  Press}, \bibinfo{address}{New York}, \bibinfo{year}{1959}{\natexlab{a}}).

\bibitem[{\citenamefont{Negele and Orland}(1988)}]{negele_orland}
\bibinfo{author}{\bibfnamefont{J.~F.} \bibnamefont{Negele}} \bibnamefont{and}
  \bibinfo{author}{\bibfnamefont{H.}~\bibnamefont{Orland}},
  \emph{\bibinfo{title}{Quantum Many--Particle Systems}}
  (\bibinfo{publisher}{Addison--Wesley}, \bibinfo{address}{New York},
  \bibinfo{year}{1988}).

\bibitem[{\citenamefont{Landau and
  Lifshitz}(1959{\natexlab{b}})}]{landau_elasticity}
\bibinfo{author}{\bibfnamefont{L.~D.} \bibnamefont{Landau}} \bibnamefont{and}
  \bibinfo{author}{\bibfnamefont{E.~M.} \bibnamefont{Lifshitz}},
  \emph{\bibinfo{title}{Theory of Elasticity}} (\bibinfo{publisher}{Pergamon
  Press}, \bibinfo{address}{New York}, \bibinfo{year}{1959}{\natexlab{b}}).

\bibitem[{\citenamefont{Zhitomirsky and
  Honecker}(2004)}]{zhitomirsky04_magnetocaloric}
\bibinfo{author}{\bibfnamefont{M.~E.} \bibnamefont{Zhitomirsky}}
  \bibnamefont{and} \bibinfo{author}{\bibfnamefont{A.}~\bibnamefont{Honecker}},
  \bibinfo{journal}{J. Stat. Mech.: Theory Exp.}
  \textbf{\bibinfo{volume}{2004}}, \bibinfo{pages}{P07012}
  (\bibinfo{year}{2004}).

\bibitem[{\citenamefont{Japaridze and
  Nersesyan}(1978)}]{japaridze_cic_transition}
\bibinfo{author}{\bibfnamefont{G.~I.} \bibnamefont{Japaridze}}
  \bibnamefont{and} \bibinfo{author}{\bibfnamefont{A.~A.}
  \bibnamefont{Nersesyan}}, \bibinfo{journal}{JETP Lett.}
  \textbf{\bibinfo{volume}{27}}, \bibinfo{pages}{334} (\bibinfo{year}{1978}).

\bibitem[{\citenamefont{Pokrovsky and Talapov}(1979)}]{pokrovsky_talapov_prl}
\bibinfo{author}{\bibfnamefont{V.~L.} \bibnamefont{Pokrovsky}}
  \bibnamefont{and} \bibinfo{author}{\bibfnamefont{A.~L.}
  \bibnamefont{Talapov}}, \bibinfo{journal}{Phys. Rev. Lett.}
  \textbf{\bibinfo{volume}{42}}, \bibinfo{pages}{65} (\bibinfo{year}{1979}).

\bibitem[{\citenamefont{Schulz}(1980)}]{schulz_cic2d}
\bibinfo{author}{\bibfnamefont{H.~J.} \bibnamefont{Schulz}},
  \bibinfo{journal}{Phys. Rev. B} \textbf{\bibinfo{volume}{22}},
  \bibinfo{pages}{5274} (\bibinfo{year}{1980}).

\bibitem[{\citenamefont{Chitra and Giamarchi}(1997)}]{chitra_spinchains_field}
\bibinfo{author}{\bibfnamefont{R.}~\bibnamefont{Chitra}} \bibnamefont{and}
  \bibinfo{author}{\bibfnamefont{T.}~\bibnamefont{Giamarchi}},
  \bibinfo{journal}{Phys. Rev. B} \textbf{\bibinfo{volume}{55}},
  \bibinfo{pages}{5816} (\bibinfo{year}{1997}).

\bibitem[{\citenamefont{Cabra and Drut}(2003)}]{cabra_instabilityLL}
\bibinfo{author}{\bibfnamefont{D.~C.} \bibnamefont{Cabra}} \bibnamefont{and}
  \bibinfo{author}{\bibfnamefont{J.~E.} \bibnamefont{Drut}},
  \bibinfo{journal}{J. Phys.: Condens. Matter} \textbf{\bibinfo{volume}{15}},
  \bibinfo{pages}{1445} (\bibinfo{year}{2003}).

\bibitem[{\citenamefont{Orignac and
  Citro}(2005)}]{orignac_2005_magnetostriction}
\bibinfo{author}{\bibfnamefont{E.}~\bibnamefont{Orignac}} \bibnamefont{and}
  \bibinfo{author}{\bibfnamefont{R.}~\bibnamefont{Citro}},
  \bibinfo{journal}{Phys. Rev. B} \textbf{\bibinfo{volume}{71}},
  \bibinfo{pages}{214419} (\bibinfo{year}{2005}). 
\bibitem[{\citenamefont{Sachdev et~al.}(1994)\citenamefont{Sachdev, Senthil,
  and Shankar}}]{sachdev_qaf_magfield}
\bibinfo{author}{\bibfnamefont{S.}~\bibnamefont{Sachdev}},
  \bibinfo{author}{\bibfnamefont{T.}~\bibnamefont{Senthil}}, \bibnamefont{and}
  \bibinfo{author}{\bibfnamefont{R.}~\bibnamefont{Shankar}},
  \bibinfo{journal}{Phys. Rev. B} \textbf{\bibinfo{volume}{50}},
  \bibinfo{pages}{258} (\bibinfo{year}{1994}).

\bibitem[{\citenamefont{Blosser et~al.}(2017)\citenamefont{Blosser, Kestin,
  Povarov, Bewley, Coira, Giamarchi, and Zheludev}}]{blosser_2017}
\bibinfo{author}{\bibfnamefont{D.}~\bibnamefont{Blosser}},
  \bibinfo{author}{\bibfnamefont{N.}~\bibnamefont{Kestin}},
  \bibinfo{author}{\bibfnamefont{K.~Y.} \bibnamefont{Povarov}},
  \bibinfo{author}{\bibfnamefont{R.}~\bibnamefont{Bewley}},
  \bibinfo{author}{\bibfnamefont{E.}~\bibnamefont{Coira}},
  \bibinfo{author}{\bibfnamefont{T.}~\bibnamefont{Giamarchi}},
  \bibnamefont{and} \bibinfo{author}{\bibfnamefont{A.}~\bibnamefont{Zheludev}},
  \bibinfo{journal}{Phys. Rev. B} \textbf{\bibinfo{volume}{96}},
  \bibinfo{pages}{134406} (\bibinfo{year}{2017}), \bibinfo{note}{arXiv:
  1707.05243}. 

\bibitem[{\citenamefont{Blosser et~al.}(2018)\citenamefont{Blosser, Bhartiya,
  Voneshen, and Zheludev}}]{blosser_2018}
\bibinfo{author}{\bibfnamefont{D.}~\bibnamefont{Blosser}},
  \bibinfo{author}{\bibfnamefont{V.~K.} \bibnamefont{Bhartiya}},
  \bibinfo{author}{\bibfnamefont{D.~J.} \bibnamefont{Voneshen}},
  \bibnamefont{and} \bibinfo{author}{\bibfnamefont{A.}~\bibnamefont{Zheludev}},
  \bibinfo{journal}{Phys. Rev. Lett.} \textbf{\bibinfo{volume}{121}},
  \bibinfo{pages}{247201} (\bibinfo{year}{2018}), \bibinfo{note}{arXiv:
  1806.10392}. 

\bibitem[{\citenamefont{Maeda et~al.}(2007)\citenamefont{Maeda, Hotta, and
  Oshikawa}}]{maeda_universal_2007}
\bibinfo{author}{\bibfnamefont{Y.}~\bibnamefont{Maeda}},
  \bibinfo{author}{\bibfnamefont{C.}~\bibnamefont{Hotta}}, \bibnamefont{and}
  \bibinfo{author}{\bibfnamefont{M.}~\bibnamefont{Oshikawa}},
  \bibinfo{journal}{Phys. Rev. Lett.} \textbf{\bibinfo{volume}{99}},
  \bibinfo{pages}{057205} (\bibinfo{year}{2007}), \bibinfo{note}{arXiv:
  cond-mat/0703727}.

\bibitem[{\citenamefont{Jordan and Wigner}(1928)}]{jordan_transformation}
\bibinfo{author}{\bibfnamefont{P.}~\bibnamefont{Jordan}} \bibnamefont{and}
  \bibinfo{author}{\bibfnamefont{E.}~\bibnamefont{Wigner}},
  \bibinfo{journal}{Z. Phys.} \textbf{\bibinfo{volume}{47}},
  \bibinfo{pages}{631} (\bibinfo{year}{1928}).

\bibitem[{\citenamefont{Katsura}(1962)}]{katsura_1962}
\bibinfo{author}{\bibfnamefont{S.}~\bibnamefont{Katsura}},
  \bibinfo{journal}{Phys. Rev.} \textbf{\bibinfo{volume}{127}},
  \bibinfo{pages}{1508} (\bibinfo{year}{1962}), \bibinfo{note}{[Erratum: Phys.
  Rev. \textbf{129}, 2835 (1963)]}.

\bibitem[{\citenamefont{Caprani}(2023)}]{caprani_absorption_2023}
\bibinfo{author}{\bibfnamefont{E.}~\bibnamefont{Caprani}},
  \bibinfo{type}{Internship report {Master 2} (in French)},
  \bibinfo{institution}{\'Ecole Normale Sup\'erieure de Lyon},
  \bibinfo{address}{Lyon, France} (\bibinfo{year}{2023}).

\bibitem[{\citenamefont{Mattis and Schultz}(1963)}]{mattis63_magnetostriction}
\bibinfo{author}{\bibfnamefont{D.~C.} \bibnamefont{Mattis}} \bibnamefont{and}
  \bibinfo{author}{\bibfnamefont{T.~D.} \bibnamefont{Schultz}},
  \bibinfo{journal}{Phys. Rev.} \textbf{\bibinfo{volume}{129}},
  \bibinfo{pages}{175} (\bibinfo{year}{1963}).

\bibitem[{\citenamefont{Derzhko et~al.}(2013)\citenamefont{Derzhko, Strečka,
  and Gálisová}}]{derzhko_2013}
\bibinfo{author}{\bibfnamefont{O.}~\bibnamefont{Derzhko}},
  \bibinfo{author}{\bibfnamefont{J.}~\bibnamefont{Strečka}}, \bibnamefont{and}
  \bibinfo{author}{\bibfnamefont{L.}~\bibnamefont{Gálisová}},
  \bibinfo{journal}{The European Physical Journal B}
  \textbf{\bibinfo{volume}{86}}, \bibinfo{pages}{88}
  (\bibinfo{year}{2013}). 

\bibitem[{\citenamefont{Olver et~al.}(2010)\citenamefont{Olver, Lozier,
  Boisvert, and Clark}}]{olver2010nist}
\bibinfo{editor}{\bibfnamefont{F.}~\bibnamefont{Olver}},
  \bibinfo{editor}{\bibfnamefont{D.}~\bibnamefont{Lozier}},
  \bibinfo{editor}{\bibfnamefont{R.}~\bibnamefont{Boisvert}}, \bibnamefont{and}
  \bibinfo{editor}{\bibfnamefont{C.}~\bibnamefont{Clark}}, eds.,
  \emph{\bibinfo{title}{NIST handbook of mathematical functions}}
  (\bibinfo{publisher}{Cambridge University Press},
  \bibinfo{address}{Cambridge, UK}, \bibinfo{year}{2010}), ISBN
  \bibinfo{isbn}{9780521140638}.

\bibitem[{\citenamefont{Guan}(2014)}]{guan_critical_2014}
\bibinfo{author}{\bibfnamefont{X.}~\bibnamefont{Guan}}, \bibinfo{journal}{Int.
  J. Mod. Phys. B} \textbf{\bibinfo{volume}{28}}, \bibinfo{pages}{1430015}
  (\bibinfo{year}{2014}), \bibinfo{note}{arXiv:1408.4473 [cond-mat]}. 

\bibitem[{\citenamefont{Zheludev}(2020)}]{zheludev_2020}
\bibinfo{author}{\bibfnamefont{A.}~\bibnamefont{Zheludev}},
  \bibinfo{journal}{Journal of Experimental and Theoretical Physics}
  \textbf{\bibinfo{volume}{131}}, \bibinfo{pages}{34} (\bibinfo{year}{2020}),
  \bibinfo{note}{arXiv: 2004.06012}. 

\bibitem[{\citenamefont{Oshikawa et~al.}(1997)\citenamefont{Oshikawa, Yamanaka,
  and Affleck}}]{oshikawa_plateaus}
\bibinfo{author}{\bibfnamefont{M.}~\bibnamefont{Oshikawa}},
  \bibinfo{author}{\bibfnamefont{M.}~\bibnamefont{Yamanaka}}, \bibnamefont{and}
  \bibinfo{author}{\bibfnamefont{I.}~\bibnamefont{Affleck}},
  \bibinfo{journal}{Phys. Rev. Lett.} \textbf{\bibinfo{volume}{78}},
  \bibinfo{pages}{1984} (\bibinfo{year}{1997}).

\bibitem[{\citenamefont{Cabra et~al.}(1998)\citenamefont{Cabra, Honecker, and
  Pujol}}]{cabra_magnetization_plateaus}
\bibinfo{author}{\bibfnamefont{D.~C.} \bibnamefont{Cabra}},
  \bibinfo{author}{\bibfnamefont{A.}~\bibnamefont{Honecker}}, \bibnamefont{and}
  \bibinfo{author}{\bibfnamefont{P.}~\bibnamefont{Pujol}},
  \bibinfo{journal}{Phys. Rev. B} \textbf{\bibinfo{volume}{58}},
  \bibinfo{pages}{6241} (\bibinfo{year}{1998}).

\bibitem[{\citenamefont{Destri and de~Vega}(1997)}]{destri97_tba_inteq}
\bibinfo{author}{\bibfnamefont{C.}~\bibnamefont{Destri}} \bibnamefont{and}
  \bibinfo{author}{\bibfnamefont{H.}~\bibnamefont{de~Vega}},
  \bibinfo{journal}{Nucl. Phys. B} \textbf{\bibinfo{volume}{504}},
  \bibinfo{pages}{621} (\bibinfo{year}{1997}).

\bibitem[{\citenamefont{He et~al.}(2017)\citenamefont{He, Jiang, Yu, Lin, and
  Guan}}]{he_2017}
\bibinfo{author}{\bibfnamefont{F.}~\bibnamefont{He}},
  \bibinfo{author}{\bibfnamefont{Y.}~\bibnamefont{Jiang}},
  \bibinfo{author}{\bibfnamefont{Y.-C.} \bibnamefont{Yu}},
  \bibinfo{author}{\bibfnamefont{H.-Q.} \bibnamefont{Lin}}, \bibnamefont{and}
  \bibinfo{author}{\bibfnamefont{X.-W.} \bibnamefont{Guan}},
  \bibinfo{journal}{Phys. Rev. B} \textbf{\bibinfo{volume}{96}},
  \bibinfo{pages}{220401} (\bibinfo{year}{2017}),
  \bibinfo{note}{arXiv:1702.05903}. 

\bibitem[{\citenamefont{Breunig et~al.}(2017)\citenamefont{Breunig, Garst,
  Klümper, Rohrkamp, Turnbull, and Lorenz}}]{breunig_quantum_2017}
\bibinfo{author}{\bibfnamefont{O.}~\bibnamefont{Breunig}},
  \bibinfo{author}{\bibfnamefont{M.}~\bibnamefont{Garst}},
  \bibinfo{author}{\bibfnamefont{A.}~\bibnamefont{Klümper}},
  \bibinfo{author}{\bibfnamefont{J.}~\bibnamefont{Rohrkamp}},
  \bibinfo{author}{\bibfnamefont{M.~M.} \bibnamefont{Turnbull}},
  \bibnamefont{and} \bibinfo{author}{\bibfnamefont{T.}~\bibnamefont{Lorenz}},
  \bibinfo{journal}{Science Advances} \textbf{\bibinfo{volume}{3}},
  \bibinfo{pages}{eaao3773} (\bibinfo{year}{2017}). 

\bibitem[{\citenamefont{Yang and Yang}(1966)}]{yang_xxz}
\bibinfo{author}{\bibfnamefont{C.~N.} \bibnamefont{Yang}} \bibnamefont{and}
  \bibinfo{author}{\bibfnamefont{C.~P.} \bibnamefont{Yang}},
  \bibinfo{journal}{Phys. Rev.} \textbf{\bibinfo{volume}{150}},
  \bibinfo{pages}{327} (\bibinfo{year}{1966}).

\bibitem[{\citenamefont{den Nijs}(1981)}]{nijs_equivalence}
\bibinfo{author}{\bibfnamefont{M.~P.~M.} \bibnamefont{den Nijs}},
  \bibinfo{journal}{Phys. Rev. B} \textbf{\bibinfo{volume}{23}},
  \bibinfo{pages}{6111} (\bibinfo{year}{1981}).

\bibitem[{\citenamefont{Haldane}(1980)}]{haldane_xxzchain}
\bibinfo{author}{\bibfnamefont{F.~D.~M.} \bibnamefont{Haldane}},
  \bibinfo{journal}{Phys. Rev. Lett.} \textbf{\bibinfo{volume}{45}},
  \bibinfo{pages}{1358} (\bibinfo{year}{1980}).

\bibitem[{\citenamefont{Mahan}(1981)}]{mahan_book}
\bibinfo{author}{\bibfnamefont{G.~D.} \bibnamefont{Mahan}},
  \emph{\bibinfo{title}{Many Particle Physics}} (\bibinfo{publisher}{Plenum},
  \bibinfo{address}{New York}, \bibinfo{year}{1981}).

\bibitem[{\citenamefont{Giamarchi and Schulz}(1989)}]{giamarchi_logs}
\bibinfo{author}{\bibfnamefont{T.}~\bibnamefont{Giamarchi}} \bibnamefont{and}
  \bibinfo{author}{\bibfnamefont{H.~J.} \bibnamefont{Schulz}},
  \bibinfo{journal}{Phys. Rev. B} \textbf{\bibinfo{volume}{39}},
  \bibinfo{pages}{4620} (\bibinfo{year}{1989}).

\bibitem[{\citenamefont{Zapf et~al.}(2014)\citenamefont{Zapf, Jaime, and
  Batista}}]{zapf_2014}
\bibinfo{author}{\bibfnamefont{V.}~\bibnamefont{Zapf}},
  \bibinfo{author}{\bibfnamefont{M.}~\bibnamefont{Jaime}}, \bibnamefont{and}
  \bibinfo{author}{\bibfnamefont{C.~D.} \bibnamefont{Batista}},
  \bibinfo{journal}{Reviews of Modern Physics} \textbf{\bibinfo{volume}{86}},
  \bibinfo{pages}{563} (\bibinfo{year}{2014}).

\bibitem[{\citenamefont{Klanjsek et~al.}(2015)\citenamefont{Klanjsek, Horvatic,
  Kramer, Mukhopadhyay, Mayaffre, Berthier, Canevet, Grenier, Lejay, and
  Orignac}}]{klanjsek_2015}
\bibinfo{author}{\bibfnamefont{M.}~\bibnamefont{Klanjsek}},
  \bibinfo{author}{\bibfnamefont{M.}~\bibnamefont{Horvatic}},
  \bibinfo{author}{\bibfnamefont{S.}~\bibnamefont{Kramer}},
  \bibinfo{author}{\bibfnamefont{S.}~\bibnamefont{Mukhopadhyay}},
  \bibinfo{author}{\bibfnamefont{H.}~\bibnamefont{Mayaffre}},
  \bibinfo{author}{\bibfnamefont{C.}~\bibnamefont{Berthier}},
  \bibinfo{author}{\bibfnamefont{E.}~\bibnamefont{Canevet}},
  \bibinfo{author}{\bibfnamefont{B.}~\bibnamefont{Grenier}},
  \bibinfo{author}{\bibfnamefont{P.}~\bibnamefont{Lejay}}, \bibnamefont{and}
  \bibinfo{author}{\bibfnamefont{E.}~\bibnamefont{Orignac}},
  \bibinfo{journal}{Phys. Rev. B} \textbf{\bibinfo{volume}{92}},
  \bibinfo{pages}{060408(R)} (\bibinfo{year}{2015}).

\bibitem[{\citenamefont{Faure et~al.}(2019)\citenamefont{Faure, Takayoshi,
  Simonet, Grenier, Månsson, White, Tucker, Rüegg, Lejay, Giamarchi
  et~al.}}]{faure_2019}
\bibinfo{author}{\bibfnamefont{Q.}~\bibnamefont{Faure}},
  \bibinfo{author}{\bibfnamefont{S.}~\bibnamefont{Takayoshi}},
  \bibinfo{author}{\bibfnamefont{V.}~\bibnamefont{Simonet}},
  \bibinfo{author}{\bibfnamefont{B.}~\bibnamefont{Grenier}},
  \bibinfo{author}{\bibfnamefont{M.}~\bibnamefont{Månsson}},
  \bibinfo{author}{\bibfnamefont{J.~S.} \bibnamefont{White}},
  \bibinfo{author}{\bibfnamefont{G.~S.} \bibnamefont{Tucker}},
  \bibinfo{author}{\bibfnamefont{C.}~\bibnamefont{Rüegg}},
  \bibinfo{author}{\bibfnamefont{P.}~\bibnamefont{Lejay}},
  \bibinfo{author}{\bibfnamefont{T.}~\bibnamefont{Giamarchi}},
  \bibnamefont{et~al.}, \bibinfo{journal}{Physical Review Letters}
  \textbf{\bibinfo{volume}{123}}, \bibinfo{pages}{027204}
  (\bibinfo{year}{2019}), \bibinfo{note}{arXiv: 1903.04173}. 

\bibitem[{\citenamefont{Zheludev et~al.}(2002)\citenamefont{Zheludev, Honda,
  Chen, Broholm, Katsumata, and Shapiro}}]{zheludev_ndmap_2002}
\bibinfo{author}{\bibfnamefont{A.}~\bibnamefont{Zheludev}},
  \bibinfo{author}{\bibfnamefont{Z.}~\bibnamefont{Honda}},
  \bibinfo{author}{\bibfnamefont{Y.}~\bibnamefont{Chen}},
  \bibinfo{author}{\bibfnamefont{C.~L.} \bibnamefont{Broholm}},
  \bibinfo{author}{\bibfnamefont{K.}~\bibnamefont{Katsumata}},
  \bibnamefont{and} \bibinfo{author}{\bibfnamefont{S.~M.}
  \bibnamefont{Shapiro}}, \bibinfo{journal}{Physical Review Letters}
  \textbf{\bibinfo{volume}{88}}, \bibinfo{pages}{077206}
  (\bibinfo{year}{2002}). 

\bibitem[{\citenamefont{Bera et~al.}(2013)\citenamefont{Bera, Lake, Islam,
  Klemke, Faulhaber, and Law}}]{bera_field_2013}
\bibinfo{author}{\bibfnamefont{A.~K.} \bibnamefont{Bera}},
  \bibinfo{author}{\bibfnamefont{B.}~\bibnamefont{Lake}},
  \bibinfo{author}{\bibfnamefont{A.~T. M.~N.} \bibnamefont{Islam}},
  \bibinfo{author}{\bibfnamefont{B.}~\bibnamefont{Klemke}},
  \bibinfo{author}{\bibfnamefont{E.}~\bibnamefont{Faulhaber}},
  \bibnamefont{and} \bibinfo{author}{\bibfnamefont{J.~M.} \bibnamefont{Law}},
  \bibinfo{journal}{Physical Review B} \textbf{\bibinfo{volume}{87}},
  \bibinfo{pages}{224423} (\bibinfo{year}{2013}),
  \bibinfo{note}{arXiv:1310.0221 [cond-mat.str-el]}. 

\bibitem[{\citenamefont{Li et~al.}(2025)\citenamefont{Li, Wu, Wang, Luo, Du,
  Xu, Hu, Chen, Yang, Liu et~al.}}]{li_haldane_2025}
\bibinfo{author}{\bibfnamefont{S.}~\bibnamefont{Li}},
  \bibinfo{author}{\bibfnamefont{Z.}~\bibnamefont{Wu}},
  \bibinfo{author}{\bibfnamefont{Y.}~\bibnamefont{Wang}},
  \bibinfo{author}{\bibfnamefont{J.}~\bibnamefont{Luo}},
  \bibinfo{author}{\bibfnamefont{K.}~\bibnamefont{Du}},
  \bibinfo{author}{\bibfnamefont{X.}~\bibnamefont{Xu}},
  \bibinfo{author}{\bibfnamefont{Z.}~\bibnamefont{Hu}},
  \bibinfo{author}{\bibfnamefont{Y.}~\bibnamefont{Chen}},
  \bibinfo{author}{\bibfnamefont{J.}~\bibnamefont{Yang}},
  \bibinfo{author}{\bibfnamefont{Z.}~\bibnamefont{Liu}}, \bibnamefont{et~al.},
  \bibinfo{journal}{Physical Review B} \textbf{\bibinfo{volume}{111}},
  \bibinfo{pages}{195164} (\bibinfo{year}{2025}),
  \bibinfo{note}{arXiv:2411.19538 [cond-mat]}.

\bibitem[{\citenamefont{Klanj\v{s}ek et~al.}(2008)\citenamefont{Klanj\v{s}ek,
  Mayaffre, Berthier, Horvati\'{c}, Chiari, Piovesana, Bouillot, Kollath,
  Orignac, Citro et~al.}}]{Klanjsek-PRL-2008}
\bibinfo{author}{\bibfnamefont{M.}~\bibnamefont{Klanj\v{s}ek}},
  \bibinfo{author}{\bibfnamefont{H.}~\bibnamefont{Mayaffre}},
  \bibinfo{author}{\bibfnamefont{C.}~\bibnamefont{Berthier}},
  \bibinfo{author}{\bibfnamefont{M.}~\bibnamefont{Horvati\'{c}}},
  \bibinfo{author}{\bibfnamefont{B.}~\bibnamefont{Chiari}},
  \bibinfo{author}{\bibfnamefont{O.}~\bibnamefont{Piovesana}},
  \bibinfo{author}{\bibfnamefont{P.}~\bibnamefont{Bouillot}},
  \bibinfo{author}{\bibfnamefont{C.}~\bibnamefont{Kollath}},
  \bibinfo{author}{\bibfnamefont{E.}~\bibnamefont{Orignac}},
  \bibinfo{author}{\bibfnamefont{R.}~\bibnamefont{Citro}},
  \bibnamefont{et~al.}, \bibinfo{journal}{Phys. Rev. Lett.}
  \textbf{\bibinfo{volume}{101}}, \bibinfo{pages}{137207}
  (\bibinfo{year}{2008}).

\bibitem[{\citenamefont{Hong et~al.}(2010)\citenamefont{Hong, Kim, Hotta,
  Takano, Tremelling, Turnbull, Landee, Kang, Christensen, Lefmann
  et~al.}}]{hong:2010}
\bibinfo{author}{\bibfnamefont{T.}~\bibnamefont{Hong}},
  \bibinfo{author}{\bibfnamefont{Y.~H.} \bibnamefont{Kim}},
  \bibinfo{author}{\bibfnamefont{C.}~\bibnamefont{Hotta}},
  \bibinfo{author}{\bibfnamefont{Y.}~\bibnamefont{Takano}},
  \bibinfo{author}{\bibfnamefont{G.}~\bibnamefont{Tremelling}},
  \bibinfo{author}{\bibfnamefont{M.~M.} \bibnamefont{Turnbull}},
  \bibinfo{author}{\bibfnamefont{C.~P.} \bibnamefont{Landee}},
  \bibinfo{author}{\bibfnamefont{H.-J.} \bibnamefont{Kang}},
  \bibinfo{author}{\bibfnamefont{N.~B.} \bibnamefont{Christensen}},
  \bibinfo{author}{\bibfnamefont{K.}~\bibnamefont{Lefmann}},
  \bibnamefont{et~al.}, \bibinfo{journal}{Phys. Rev. Lett.}
  \textbf{\bibinfo{volume}{105}}, \bibinfo{pages}{137207}
  (\bibinfo{year}{2010}).

\bibitem[{\citenamefont{Chen et~al.}(2024)\citenamefont{Chen, Hu, Qu, Li, Liu,
  Wang, Sun, Dong, and Qiu}}]{chen_zigzag_2024}
\bibinfo{author}{\bibfnamefont{R.}~\bibnamefont{Chen}},
  \bibinfo{author}{\bibfnamefont{H.~J.} \bibnamefont{Hu}},
  \bibinfo{author}{\bibfnamefont{Z.}~\bibnamefont{Qu}},
  \bibinfo{author}{\bibfnamefont{T.}~\bibnamefont{Li}},
  \bibinfo{author}{\bibfnamefont{C.~B.} \bibnamefont{Liu}},
  \bibinfo{author}{\bibfnamefont{C.~L.} \bibnamefont{Wang}},
  \bibinfo{author}{\bibfnamefont{S.~J.} \bibnamefont{Sun}},
  \bibinfo{author}{\bibfnamefont{C.}~\bibnamefont{Dong}}, \bibnamefont{and}
  \bibinfo{author}{\bibfnamefont{Y.}~\bibnamefont{Qiu}},
  \bibinfo{journal}{Journal of Physics: Condensed Matter}
  \textbf{\bibinfo{volume}{36}}, \bibinfo{pages}{165801}
  (\bibinfo{year}{2024}).

\bibitem[{\citenamefont{Yoshida et~al.}(2005)\citenamefont{Yoshida, Tateiwa,
  Mito, Kawae, Takeda, Hosokoshi, and Inoue}}]{yoshida_f5pnn_2005}
\bibinfo{author}{\bibfnamefont{Y.}~\bibnamefont{Yoshida}},
  \bibinfo{author}{\bibfnamefont{N.}~\bibnamefont{Tateiwa}},
  \bibinfo{author}{\bibfnamefont{M.}~\bibnamefont{Mito}},
  \bibinfo{author}{\bibfnamefont{T.}~\bibnamefont{Kawae}},
  \bibinfo{author}{\bibfnamefont{K.}~\bibnamefont{Takeda}},
  \bibinfo{author}{\bibfnamefont{Y.}~\bibnamefont{Hosokoshi}},
  \bibnamefont{and} \bibinfo{author}{\bibfnamefont{K.}~\bibnamefont{Inoue}},
  \bibinfo{journal}{Phys. Rev. Lett.} \textbf{\bibinfo{volume}{94}},
  \bibinfo{pages}{037203} (\bibinfo{year}{2005}).

\bibitem[{\citenamefont{Stone et~al.}(2014)\citenamefont{Stone, Chen, Reich,
  Broholm, Xu, Copley, and Cook}}]{stone_cunitrate_2014}
\bibinfo{author}{\bibfnamefont{M.~B.} \bibnamefont{Stone}},
  \bibinfo{author}{\bibfnamefont{Y.}~\bibnamefont{Chen}},
  \bibinfo{author}{\bibfnamefont{D.~H.} \bibnamefont{Reich}},
  \bibinfo{author}{\bibfnamefont{C.}~\bibnamefont{Broholm}},
  \bibinfo{author}{\bibfnamefont{G.}~\bibnamefont{Xu}},
  \bibinfo{author}{\bibfnamefont{J.~R.~D.} \bibnamefont{Copley}},
  \bibnamefont{and} \bibinfo{author}{\bibfnamefont{J.~C.} \bibnamefont{Cook}},
  \bibinfo{journal}{Phys. Rev. B} \textbf{\bibinfo{volume}{90}},
  \bibinfo{pages}{094419} (\bibinfo{year}{2014}),
  \bibinfo{note}{arXiv:1406.7596 [cond-mat]}.

\bibitem[{\citenamefont{Willenberg et~al.}(2015)\citenamefont{Willenberg, Ryll,
  Kiefer, Tennant, Groitl, Rolfs, Manuel, Khalyavin, Rule, Wolter
  et~al.}}]{willenberg_2015}
\bibinfo{author}{\bibfnamefont{B.}~\bibnamefont{Willenberg}},
  \bibinfo{author}{\bibfnamefont{H.}~\bibnamefont{Ryll}},
  \bibinfo{author}{\bibfnamefont{K.}~\bibnamefont{Kiefer}},
  \bibinfo{author}{\bibfnamefont{D.~A.} \bibnamefont{Tennant}},
  \bibinfo{author}{\bibfnamefont{F.}~\bibnamefont{Groitl}},
  \bibinfo{author}{\bibfnamefont{K.}~\bibnamefont{Rolfs}},
  \bibinfo{author}{\bibfnamefont{P.}~\bibnamefont{Manuel}},
  \bibinfo{author}{\bibfnamefont{D.}~\bibnamefont{Khalyavin}},
  \bibinfo{author}{\bibfnamefont{K.~C.} \bibnamefont{Rule}},
  \bibinfo{author}{\bibfnamefont{A.~U.~B.} \bibnamefont{Wolter}},
  \bibnamefont{et~al.}, \bibinfo{journal}{Phys. Rev. B}
  \textbf{\bibinfo{volume}{91}}, \bibinfo{pages}{060407}
  (\bibinfo{year}{2015}), \bibinfo{note}{arXiv:1406.6149 [cond-mat]}.

\bibitem[{\citenamefont{Islam et~al.}(2024)\citenamefont{Islam, Mukharjee,
  Biswas, Telling, Skourski, Ranjith, Baenitz, Inagaki, Furukawa, Tsirlin
  et~al.}}]{islam_2024}
\bibinfo{author}{\bibfnamefont{S.~S.} \bibnamefont{Islam}},
  \bibinfo{author}{\bibfnamefont{P.~K.} \bibnamefont{Mukharjee}},
  \bibinfo{author}{\bibfnamefont{P.~K.} \bibnamefont{Biswas}},
  \bibinfo{author}{\bibfnamefont{M.}~\bibnamefont{Telling}},
  \bibinfo{author}{\bibfnamefont{Y.}~\bibnamefont{Skourski}},
  \bibinfo{author}{\bibfnamefont{K.~M.} \bibnamefont{Ranjith}},
  \bibinfo{author}{\bibfnamefont{M.}~\bibnamefont{Baenitz}},
  \bibinfo{author}{\bibfnamefont{Y.}~\bibnamefont{Inagaki}},
  \bibinfo{author}{\bibfnamefont{Y.}~\bibnamefont{Furukawa}},
  \bibinfo{author}{\bibfnamefont{A.~A.} \bibnamefont{Tsirlin}},
  \bibnamefont{et~al.}, \bibinfo{journal}{Physical Review B}
  \textbf{\bibinfo{volume}{109}}, \bibinfo{pages}{L060406}
  (\bibinfo{year}{2024}).

\bibitem[{\citenamefont{Pfeuty}(1970)}]{pfeuty_transverse_ising}
\bibinfo{author}{\bibfnamefont{P.}~\bibnamefont{Pfeuty}},
  \bibinfo{journal}{Ann. Phys. (N. Y.)} \textbf{\bibinfo{volume}{27}},
  \bibinfo{pages}{79} (\bibinfo{year}{1970}).

\bibitem[{\citenamefont{Dutta}(2015)}]{dutta_book_2015}
\bibinfo{author}{\bibfnamefont{A.}~\bibnamefont{Dutta}},
  \emph{\bibinfo{title}{Quantum phase transitions in transverse field spin
  models: from statistical physics to quantum information}}
  (\bibinfo{publisher}{Cambridge University Press},
  \bibinfo{address}{Cambridge, UK}, \bibinfo{year}{2015}), ISBN
  \bibinfo{isbn}{978-1-107-06879-7}, \bibinfo{note}{arXiv:1012.0653
  [cond-mat.stat-mech]}.

\bibitem[{\citenamefont{Matsuura et~al.}(2020)\citenamefont{Matsuura, Cong,
  Zherlitsyn, Wosnitza, Abe, and Arima}}]{matsuura_conb2o6_2020}
\bibinfo{author}{\bibfnamefont{K.}~\bibnamefont{Matsuura}},
  \bibinfo{author}{\bibfnamefont{P.~T.} \bibnamefont{Cong}},
  \bibinfo{author}{\bibfnamefont{S.}~\bibnamefont{Zherlitsyn}},
  \bibinfo{author}{\bibfnamefont{J.}~\bibnamefont{Wosnitza}},
  \bibinfo{author}{\bibfnamefont{N.}~\bibnamefont{Abe}}, \bibnamefont{and}
  \bibinfo{author}{\bibfnamefont{T.-h.} \bibnamefont{Arima}},
  \bibinfo{journal}{Physical Review Letters} \textbf{\bibinfo{volume}{124}},
  \bibinfo{pages}{127205} (\bibinfo{year}{2020}).

\bibitem[{\citenamefont{Cl\'emancey et~al.}(2006)\citenamefont{Cl\'emancey,
  Mayaffre, Berthier, Horvatic, Fouet, Miyahara, Mila, Chiari, and
  Piovesana}}]{clemancey06_cuhpcl}
\bibinfo{author}{\bibfnamefont{M.}~\bibnamefont{Cl\'emancey}},
  \bibinfo{author}{\bibfnamefont{H.}~\bibnamefont{Mayaffre}},
  \bibinfo{author}{\bibfnamefont{C.}~\bibnamefont{Berthier}},
  \bibinfo{author}{\bibfnamefont{M.}~\bibnamefont{Horvatic}},
  \bibinfo{author}{\bibfnamefont{J.-B.} \bibnamefont{Fouet}},
  \bibinfo{author}{\bibfnamefont{S.}~\bibnamefont{Miyahara}},
  \bibinfo{author}{\bibfnamefont{F.}~\bibnamefont{Mila}},
  \bibinfo{author}{\bibfnamefont{B.}~\bibnamefont{Chiari}}, \bibnamefont{and}
  \bibinfo{author}{\bibfnamefont{O.}~\bibnamefont{Piovesana}},
  \bibinfo{journal}{Phys. Rev. Lett.} \textbf{\bibinfo{volume}{97}},
  \bibinfo{pages}{167204} (\bibinfo{year}{2006}).

\bibitem[{\citenamefont{Dzyaloshinskii}(1958)}]{dzyalo_interaction}
\bibinfo{author}{\bibfnamefont{I.}~\bibnamefont{Dzyaloshinskii}},
  \bibinfo{journal}{J. Phys. Chem. Solids} \textbf{\bibinfo{volume}{4}},
  \bibinfo{pages}{241} (\bibinfo{year}{1958}).

\bibitem[{\citenamefont{Moriya}(1960)}]{moriya60}
\bibinfo{author}{\bibfnamefont{T.}~\bibnamefont{Moriya}},
  \bibinfo{journal}{Phys. Rev.} \textbf{\bibinfo{volume}{120}},
  \bibinfo{pages}{91} (\bibinfo{year}{1960}).

\bibitem[{\citenamefont{Orignac et~al.}(2007)\citenamefont{Orignac, Citro,
  Capponi, and Poilblanc}}]{orignac07_staggered_dm}
\bibinfo{author}{\bibfnamefont{E.}~\bibnamefont{Orignac}},
  \bibinfo{author}{\bibfnamefont{R.}~\bibnamefont{Citro}},
  \bibinfo{author}{\bibfnamefont{S.}~\bibnamefont{Capponi}}, \bibnamefont{and}
  \bibinfo{author}{\bibfnamefont{D.}~\bibnamefont{Poilblanc}},
  \bibinfo{journal}{Phys. Rev. B} \textbf{\bibinfo{volume}{76}},
  \bibinfo{pages}{144422} (\bibinfo{year}{2007}), \eprint{arXiv:0706.3590}.

\bibitem[{\citenamefont{Giamarchi and
  Tsvelik}(1999)}]{giamarchi_coupled_ladders}
\bibinfo{author}{\bibfnamefont{T.}~\bibnamefont{Giamarchi}} \bibnamefont{and}
  \bibinfo{author}{\bibfnamefont{A.~M.} \bibnamefont{Tsvelik}},
  \bibinfo{journal}{Phys. Rev. B} \textbf{\bibinfo{volume}{59}},
  \bibinfo{pages}{11398} (\bibinfo{year}{1999}),
  \bibinfo{note}{cond-mat/9810219}.

\bibitem[{\citenamefont{Irkhin and Katanin}(2000)}]{irkhin00_chain_mf}
\bibinfo{author}{\bibfnamefont{V.~Y.} \bibnamefont{Irkhin}} \bibnamefont{and}
  \bibinfo{author}{\bibfnamefont{A.~A.} \bibnamefont{Katanin}},
  \bibinfo{journal}{Phys. Rev. B} \textbf{\bibinfo{volume}{61}},
  \bibinfo{pages}{6757} (\bibinfo{year}{2000}).

\bibitem[{\citenamefont{Pincus}(1971)}]{pincus_spin-peierls}
\bibinfo{author}{\bibfnamefont{P.}~\bibnamefont{Pincus}},
  \bibinfo{journal}{Solid State Commun.} \textbf{\bibinfo{volume}{4}},
  \bibinfo{pages}{1971} (\bibinfo{year}{1971}).

\bibitem[{\citenamefont{Pytte}(1974)}]{pytte_sp}
\bibinfo{author}{\bibfnamefont{E.}~\bibnamefont{Pytte}},
  \bibinfo{journal}{Phys. Rev. B} \textbf{\bibinfo{volume}{10}},
  \bibinfo{pages}{4637} (\bibinfo{year}{1974}).

\bibitem[{\citenamefont{Cross and Fisher}(1979)}]{cross_spinpeierls}
\bibinfo{author}{\bibfnamefont{M.~C.} \bibnamefont{Cross}} \bibnamefont{and}
  \bibinfo{author}{\bibfnamefont{D.~S.} \bibnamefont{Fisher}},
  \bibinfo{journal}{Phys. Rev. B} \textbf{\bibinfo{volume}{19}},
  \bibinfo{pages}{402} (\bibinfo{year}{1979}).

\bibitem[{\citenamefont{Inagaki and Fukuyama}(1983)}]{inagaki_spinpeierls}
\bibinfo{author}{\bibfnamefont{S.}~\bibnamefont{Inagaki}} \bibnamefont{and}
  \bibinfo{author}{\bibfnamefont{H.}~\bibnamefont{Fukuyama}},
  \bibinfo{journal}{J. Phys. Soc. Jpn.} \textbf{\bibinfo{volume}{52}},
  \bibinfo{pages}{3620} (\bibinfo{year}{1983}).

\bibitem[{\citenamefont{Orignac and Chitra}(2004)}]{orignac04_tbamf}
\bibinfo{author}{\bibfnamefont{E.}~\bibnamefont{Orignac}} \bibnamefont{and}
  \bibinfo{author}{\bibfnamefont{R.}~\bibnamefont{Chitra}},
  \bibinfo{journal}{Phys. Rev. B} \textbf{\bibinfo{volume}{70}},
  \bibinfo{pages}{214436} (\bibinfo{year}{2004}), \eprint{cond-mat/0407561}.

\bibitem[{\citenamefont{Citro et~al.}(2005{\natexlab{b}})\citenamefont{Citro,
  Orignac, and Giamarchi}}]{citro_2005}
\bibinfo{author}{\bibfnamefont{R.}~\bibnamefont{Citro}},
  \bibinfo{author}{\bibfnamefont{E.}~\bibnamefont{Orignac}}, \bibnamefont{and}
  \bibinfo{author}{\bibfnamefont{T.}~\bibnamefont{Giamarchi}},
  \bibinfo{journal}{Phys. Rev. B} \textbf{\bibinfo{volume}{72}},
  \bibinfo{pages}{024434} (\bibinfo{year}{2005}{\natexlab{b}}).

\bibitem[{\citenamefont{Schulz}(1996)}]{schulz_coupled_spinchains}
\bibinfo{author}{\bibfnamefont{H.~J.} \bibnamefont{Schulz}},
  \bibinfo{journal}{Phys. Rev. Lett.} \textbf{\bibinfo{volume}{77}},
  \bibinfo{pages}{2790} (\bibinfo{year}{1996}).

\bibitem[{\citenamefont{Bocquet}(2002)}]{bocquet02_chain_rpa}
\bibinfo{author}{\bibfnamefont{M.}~\bibnamefont{Bocquet}},
  \bibinfo{journal}{Phys. Rev. B} \textbf{\bibinfo{volume}{65}},
  \bibinfo{pages}{184415} (\bibinfo{year}{2002}).

\bibitem[{\citenamefont{Dupont et~al.}(2018)\citenamefont{Dupont, Capponi,
  Laflorencie, and Orignac}}]{dupont2018}
\bibinfo{author}{\bibfnamefont{M.}~\bibnamefont{Dupont}},
  \bibinfo{author}{\bibfnamefont{S.}~\bibnamefont{Capponi}},
  \bibinfo{author}{\bibfnamefont{N.}~\bibnamefont{Laflorencie}},
  \bibnamefont{and} \bibinfo{author}{\bibfnamefont{E.}~\bibnamefont{Orignac}},
  \bibinfo{journal}{Phys. Rev. B} \textbf{\bibinfo{volume}{98}},
  \bibinfo{pages}{094403} (\bibinfo{year}{2018}), \eprint{arXiv:1806.04913}.

\bibitem[{\citenamefont{Abrikosov et~al.}(1963)\citenamefont{Abrikosov, Gorkov,
  and Dzyaloshinski}}]{abrikosov_book}
\bibinfo{author}{\bibfnamefont{A.~A.} \bibnamefont{Abrikosov}},
  \bibinfo{author}{\bibfnamefont{L.~P.} \bibnamefont{Gorkov}},
  \bibnamefont{and} \bibinfo{author}{\bibfnamefont{I.~E.}
  \bibnamefont{Dzyaloshinski}}, \emph{\bibinfo{title}{Methods of Quantum Field
  Theory in Statistical Physics}} (\bibinfo{publisher}{Dover},
  \bibinfo{address}{New York}, \bibinfo{year}{1963}).

\bibitem[{\citenamefont{Press et~al.}(1992)\citenamefont{Press, Flannery,
  Teukolsky, and Vetterling}}]{press92_nr_nystrom}
\bibinfo{author}{\bibfnamefont{W.~H.} \bibnamefont{Press}},
  \bibinfo{author}{\bibfnamefont{B.~P.} \bibnamefont{Flannery}},
  \bibinfo{author}{\bibfnamefont{S.~A.} \bibnamefont{Teukolsky}},
  \bibnamefont{and} \bibinfo{author}{\bibfnamefont{W.~T.}
  \bibnamefont{Vetterling}}, \emph{\bibinfo{title}{Numerical Recipes in
  Fortran}} (\bibinfo{publisher}{Cambridge University Press},
  \bibinfo{year}{1992}), chap.~\bibinfo{chapter}{18}.

\bibitem[{\citenamefont{Fong et~al.}(1993)\citenamefont{Fong, Jefferson,
  Suyehiro, and Walton}}]{fong_guide_1993}
\bibinfo{author}{\bibfnamefont{K.~W.} \bibnamefont{Fong}},
  \bibinfo{author}{\bibfnamefont{T.~H.} \bibnamefont{Jefferson}},
  \bibinfo{author}{\bibfnamefont{T.}~\bibnamefont{Suyehiro}}, \bibnamefont{and}
  \bibinfo{author}{\bibfnamefont{L.}~\bibnamefont{Walton}},
  \emph{\bibinfo{title}{Guide to the {SLATEC} {Common} {Mathematical}
  {Library}}} (\bibinfo{year}{1993}),
  \urlprefix\url{http://www.netlib.org/slatec/}.

\end{thebibliography}

\end{document}